\documentclass[useAMS,usenatbib]{mn2e}

\usepackage{natbib}
\bibpunct{(}{)}{;}{a}{,}{,}

\usepackage{graphicx}
\usepackage{helvet}
\usepackage{natbib}
\usepackage{setspace}
\usepackage{amssymb}
\usepackage{longtable}

\title[Pre-main sequence variable stars]
{Pre-main sequence variable stars in young open cluster NGC 1893}

\author[Sneh Lata et al.]
       {Sneh Lata$^1$\thanks{E-mail: sneh@aries.res.in}, A. K. Pandey$^1$, W. P. Chen$^2$, G. Maheswar$^1$ and Neelam Chauhan$^2$\\ 
       $^1$Aryabhatta Research Institute of Observational Sciences, Manora Peak, Nainital 263129, Uttarakhand, India \\
       $^2$Institute of Astronomy, National Central University, Chung-Li 32054, Taiwan}
\date{Accepted ---------.
      Received ---------;
      }

\pagerange{\pageref{firstpage}--\pageref{lastpage}}

\def\LaTeX{L\kern-.36em\raise.3ex\hbox{a}\kern-.15em
    T\kern-.1667em\lower.7ex\hbox{E}\kern-.125emX}

\begin{document}

\label{firstpage}

\maketitle

\label{firstpage}
\begin{abstract}
We present results of multi-epoch (fourteen nights during 2007-2010) $V$-band photometry 
of the cluster NGC 1893 region to identify photometric variable stars in the cluster. 
The study identified a total of 53 stars showing photometric variability.  
The members associated with the region are identified on the basis of spectral energy distribution, $J-H/H-K$ two colour diagram and $V/V-I$
colour-magnitude diagram. The ages and masses of the majority of pre-main-sequence sources are found to be $\lesssim$ 5 Myr and in the range 0.5 $\lesssim$ $M/M_{\odot}$ $\lesssim$ 4, respectively. These pre-main-sequence sources hence could be T Tauri stars.
We also determined the physical parameters like disk mass and accretion rate from the spectral energy distribution of these T Tauri stars. 
The periods of majority of the T Tauri stars range from 0.1 to 20 day.
The brightness of Classical T Tauri stars is found to vary with larger amplitude in comparison to
Weak line T Tauri stars. It is found that the amplitude decreases with increase in mass, which could be due to the dispersal of
disks of massive stars.  

\end{abstract}

\begin {keywords} 
Open  cluster:  NGC 1893  --
colour--magnitude diagram: Variables-pre-main sequence stars
\end {keywords}

\section{Introduction}
It is now well established that circumstellar disks are an integral part of star
formation and are potential sites for planet formation (e.g. Hillenbrand 2002).
Young star clusters have significant number of pre-main-sequence (PMS) stars
with circumstellar disk and are unique laboratories to study the evolution
of disks of PMS stars.
PMS objects are generally classified into T Tauri stars (TTSs) and Herbig 
Ae/Be stars. The TTSs have masses $\lesssim$ 3 M$_{\odot}$ which are contracting
towards the main-sequence (MS), whereas Herbig Ae/Be stars have masses in the
range of $\sim$ 3-10 M$_{\odot}$. These stars are contracting towards MS or
just reached the MS. 
On the basis of the strength of the H$\alpha$ emissions, the TTSs are
further classified as Weak line TTSs (WTTSs; equivalent width (EW) $\le$ 10 \AA) or
Classical TTSs (CTTSs; EW$>$10 \AA).
Both WTTSs and CTTSs show variation in their brightness. These variations are found to occur at all wavelengths, from X-ray to infrared.
Variability time scale of TTSs ranges from few minutes to years (Appenzeller \& Mundt 1989).
The photometric variations are believed to originate from several mechanisms like rotation of a star with an asymmetrical distribution of cool spots, variable hot spots or obscuration by circumstellar dust (see Herbst et al. 1994 and reference therein). 
The Herbig Ae/Be stars also show variability
as they move across the instability region in the Hertzsprung-Russell (HR) diagram on their way to the MS.
Several systematic studies of TTSs have been carried out which revealed different type of variabilities (Herbst et al. 1994). It is now well known that
some of them show periodic variability (e.g., Hillenbrand 2002; Schaefer 1983; Bouvier et al. 1993; Bouvier 1994; Percy et al. 2006, 2010).

NGC 1893 ($l=173^{o}.58$ and $b=-01^{o}.68$) is a young cluster
(age $\sim$ 4 Myr) having a significant population of PMS sources (Sharma
et al. 2007).
NGC 1893, located at a distance of 3.25 kpc (Sharma et al. 2007) contains a number of early
type stars, a bright diffuse nebulosity and two pennant nebulae, Sim 
129 and Sim 130 (Gaze \& Shajn 1995). The region also contains at least five O-type 
stars (Hiltner 1966).
 Efforts have been made by a number of authors to identify low mass PMS stars towards the cluster NGC 1893. On the basis of optical 
and near infra-red (NIR) photometry, Vallenari et al. (1999) suggested that there could be a significant number of PMS stars 
 in the cluster region with a large spread in their ages. Using the $uvby\beta$ CCD photometry of the region Marco et al. (2001) identified $\sim50$ likely members (spectral   
type B9-A0 ) in the cluster. 
Marco et al. (2001) also suggested the existence of a significant PMS population in the cluster. In a later study, based on low resolution 
spectroscopy, Marco \& Negueruela (2002) confirmed the PMS nature of a number of sources.
Maheswar et al. (2007) and Negueruela et al. (2007) identified 
additional numbers of emission-line and NIR excess sources. A significant number of these sources are found 
concentrated towards regions near Sim 129 and Sim 130.
Maheswar et al. (2007) and Sharma et al. (2007) gave evidence for triggered star formation in the region.
Recently on the basis of optical, NIR, mid-infra red (MIR) and X-ray data Prisinzano et al. (2011) have identified 1034 and 442 Class II and Class III sources, respectively, in the region.

In this study, we present results of our multi-epoch photometric monitoring of the region containing 
NGC 1893. The observations in the $V$ band have been carried out on 14 nights from December 2007 to October 2010 in order to identify and characterize the variable stars in the NGC 1893 region. In Section 2 we discuss the observations, data reduction procedure and the spatial distribution of variables. 
Section 3 deals with the membership based on the spectral energy distribution, $V/V-I$ colour-magnitude
diagram (CMD) and $J-H/H-K$ two colour diagram (TCD). 
Section 4, 5 and 6 describe period determination, physical state of variables and characteristics of  variables. 
Finally, we conclude our paper 
with a summary of the main results obtained in the current study in Section 7.

\begin{figure*}
\includegraphics[width=12cm]{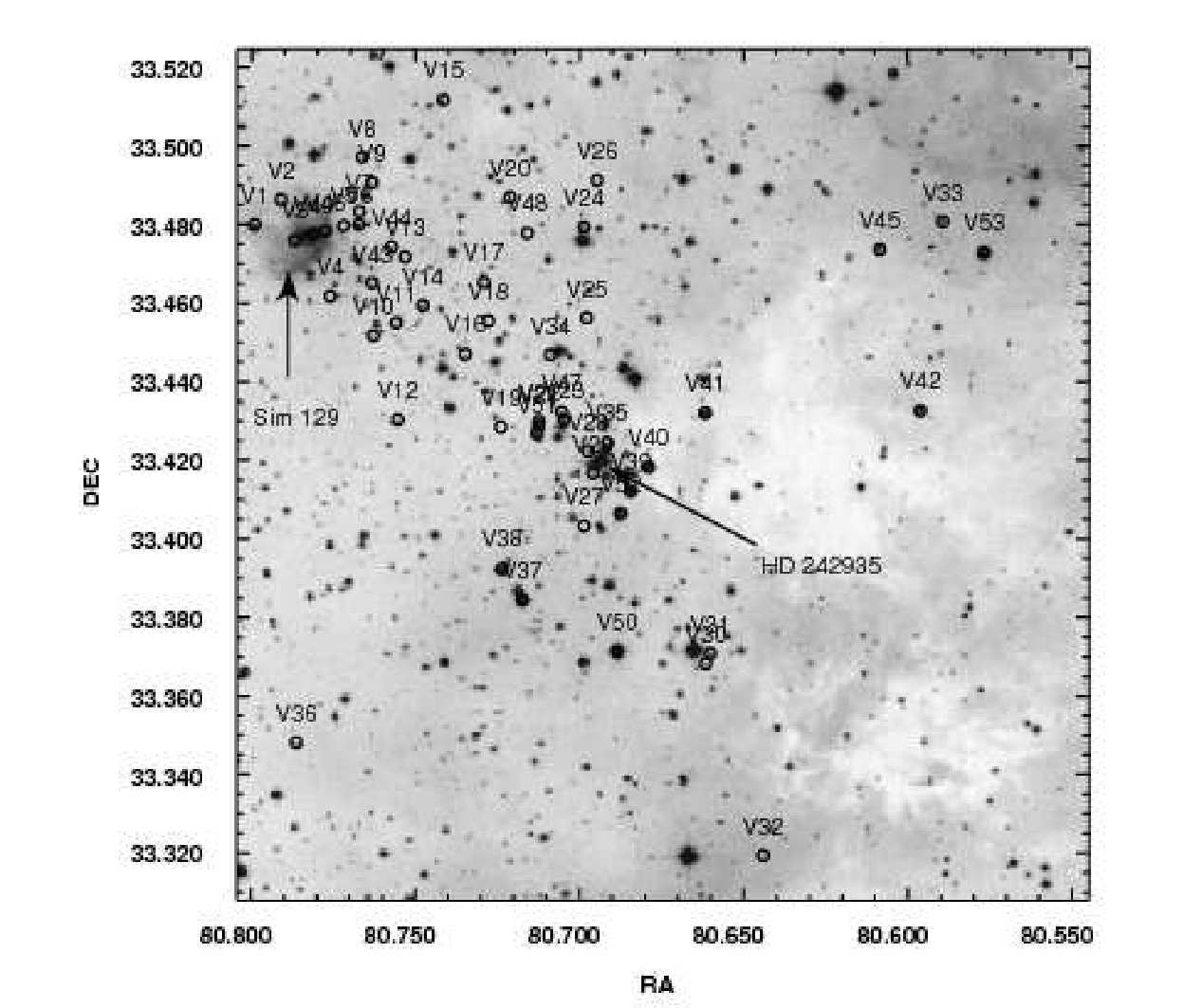}
\caption{ The observed region of NGC 1893 (image taken from the DSS-R). The circles show the location of variables identified in the present work.}
\end{figure*}


\begin{table}
\caption{Log of the observations. N and Exp. represent number of frames obtained and exposure time respectively. \label{tab:obsLog}}
\begin{tabular}{lclcc}
\hline
S. No.&Date of        &Object&{\it V}                &{\it I}                               \\
         &observations&         &(N$\times$Exp.)&(N$\times$Exp.) \\
\hline
1 & 05 Dec 2007 & NGC 1893&3$\times$40s &- \\
2 & 08 Dec 2007 & NGC 1893&3$\times$50s &-\\
3 & 07 Jan 2008 & NGC 1893&2$\times$40s &-\\
4 & 10 Jan 2008 & NGC 1893&3$\times$50s &-\\
5 & 12 Jan 2008 & NGC 1893&80$\times$50s &-\\
6 & 14 Jan 2008 & NGC 1893&70$\times$40s &-\\
7 & 29 Oct 2008 & NGC 1893&97$\times$50s &-\\
8 & 21 Nov 2008 & NGC 1893&137$\times$50s &2$\times$50s\\
9 & 27 Jan 2009 & NGC 1893&5$\times$50s &-\\
10& 28 Jan 2009 & NGC 1893&5$\times$50s &-\\
11& 19 Feb 2009 & NGC 1893&5$\times$50s &-\\
12& 20 Feb 2009 & NGC 1893&3$\times$50s &5$\times$50s\\
13& 20 Feb 2009 & SA 98     &5$\times$90s &5$\times$60s\\
14& 31 Oct 2010 & NGC 1893&3$\times$50s &-\\
\hline
\end{tabular}

\end{table}

\section{Observations and Data Reduction}
The photometric monitoring of the NGC 1893 region was carried out in the 
$V$-band on 14 nights and in the $I$-band on two nights from   
2007 December 05 to 2010 October 31 using a 2048$\times$2048 CCD camera 
attached to the 104 cm Sampurnanand ARIES telescope. The field of view is $\sim$13$^{\prime}\times$13$^{\prime}$ 
and the scale is $\sim0.76^{\prime\prime}$/pixel in 2$\times$2 pixel binning mode. The central position on 
the sky was close to RA (2000) = $05^{h}22^{m}42^{s}$ and Dec (2000) = $+33^{\circ}25^{\prime}00^{\prime\prime}$ 
for all the frames. On each night, at least two frames of target field were obtained in the $V$-band to cover 
long period variables. 
Fig. 1, taken
from Digital Sky Survey (DSS), displays the observed region of NGC 1893. 
The observations of NGC 1893 consist of a total of 421 CCD images in the $V$-band. 
The typical seeing (estimated from the FWHM of the point like stars) of the images was found to be $1.5^{\prime\prime}-2^{\prime\prime}$. Bias and twilight flats were also taken 
along with the target field. The log of the observations is given in Table \ref{tab:obsLog}. 
\subsection{Photometry}
The preprocessing of the CCD images was performed by using the IRAF\footnote{IRAF is distributed by the National Optical Astronomy Observatory, which is operated by the Association of Universities for Research in Astronomy (AURA) under cooperative agreement with the National Science Foundation.},   
which includes bias subtraction, flat field correction and removal of cosmic rays.  
The instrumental magnitude of the stars were obtained using the DAOPHOT package (Stetson 1987). Both aperture and PSF photometry were carried out to get the magnitude of the 
stars. The PSF photometry yields better results for crowded regions.
 We have used the DAOMATCH (Stetson 1992) routine of DAOPHOT  to find the translation, rotation and scaling solutions between different
photometry files, whereas DAOMASTER (Stetson 1992) matches the point sources.
To remove frame-to-frame flux variation
due to airmass and exposure time, we used DAOMASTER programme
to get the corrected magnitude.
This task makes the mean flux level of each frame equal to the
reference frame by an additive constant.
The first target frame taken on 21 November 2008 has been considered as the
the reference frame.
Present observations have 421 frames and each frame corresponds to one photometry file.
The DAOMASTER cross identified 1186 stars in different photometry files and listed their
corrected magnitudes in a .cor file.
We have considered only those stars for further study which have at least 100 observations.
The data file (.cor) generated by DAOMASTER programme was used to identify variable candidates in the next section.

The standardization of the cluster fields was carried out by observing the standard field SA 98 (Landolt 1992). The standardization procedure is outlined 
below. 
First, we have standardised the cluster region using the observations of the region and standard field SA 98 taken on 20 Feb 2009. 
The instrumental magnitudes were transformed to the standard Johnson V and Cousins I system using the procedure outlined by Stetson (1992). The equations used for photometric calibration are given below: 
\begin{eqnarray}
v = V + a_{1} - b_{1}\times (V-I) + 0.26\times Q   \nonumber\\
i = I+a_{2}  - b_{2}\times (V-I) + 0.16\times Q   \nonumber
\end{eqnarray}
where $v$ and $i$ are the  instrumental magnitudes and $Q$ is the airmass. The values of $a_{1}$, $a_{2}$, $b_{1}$, and $b_{2}$ are $5.009\pm 0.007$, $5.357\pm 0.012$, $0.0496\pm 0.006$, and $0.0575\pm 0.010$ respectively.
Stars spread  all over the cluster region were selected as secondary standards. The secondary standards were used to standardise the reference frame.


\begin{figure}
\includegraphics[width=8cm]{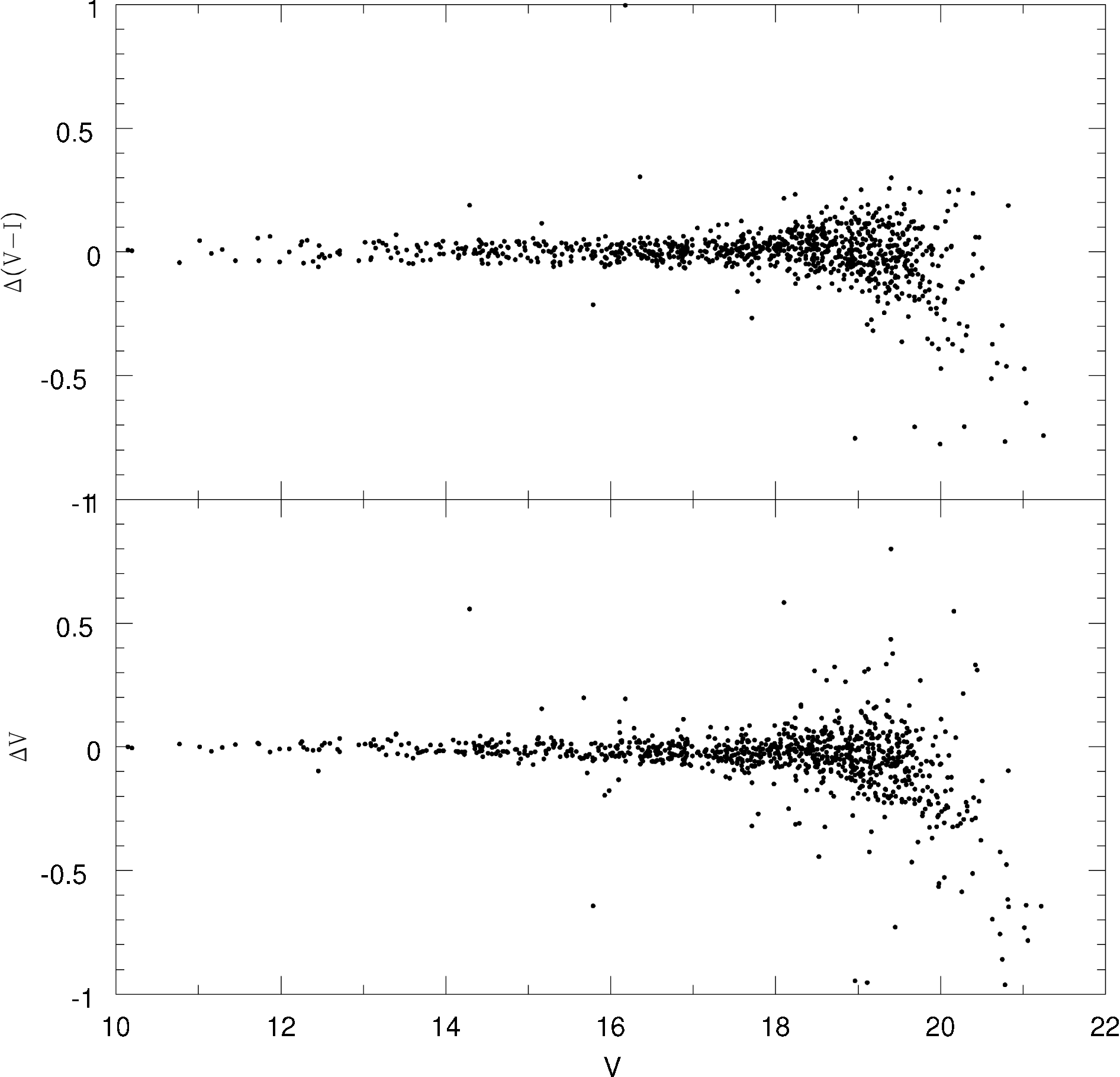}
\caption{Comparison of the present photometry and photometry given by Sharma et al. (2007). The $\Delta$ represents present minus previous photometry.}
\end{figure}

\begin{figure}
\includegraphics[width=8cm]{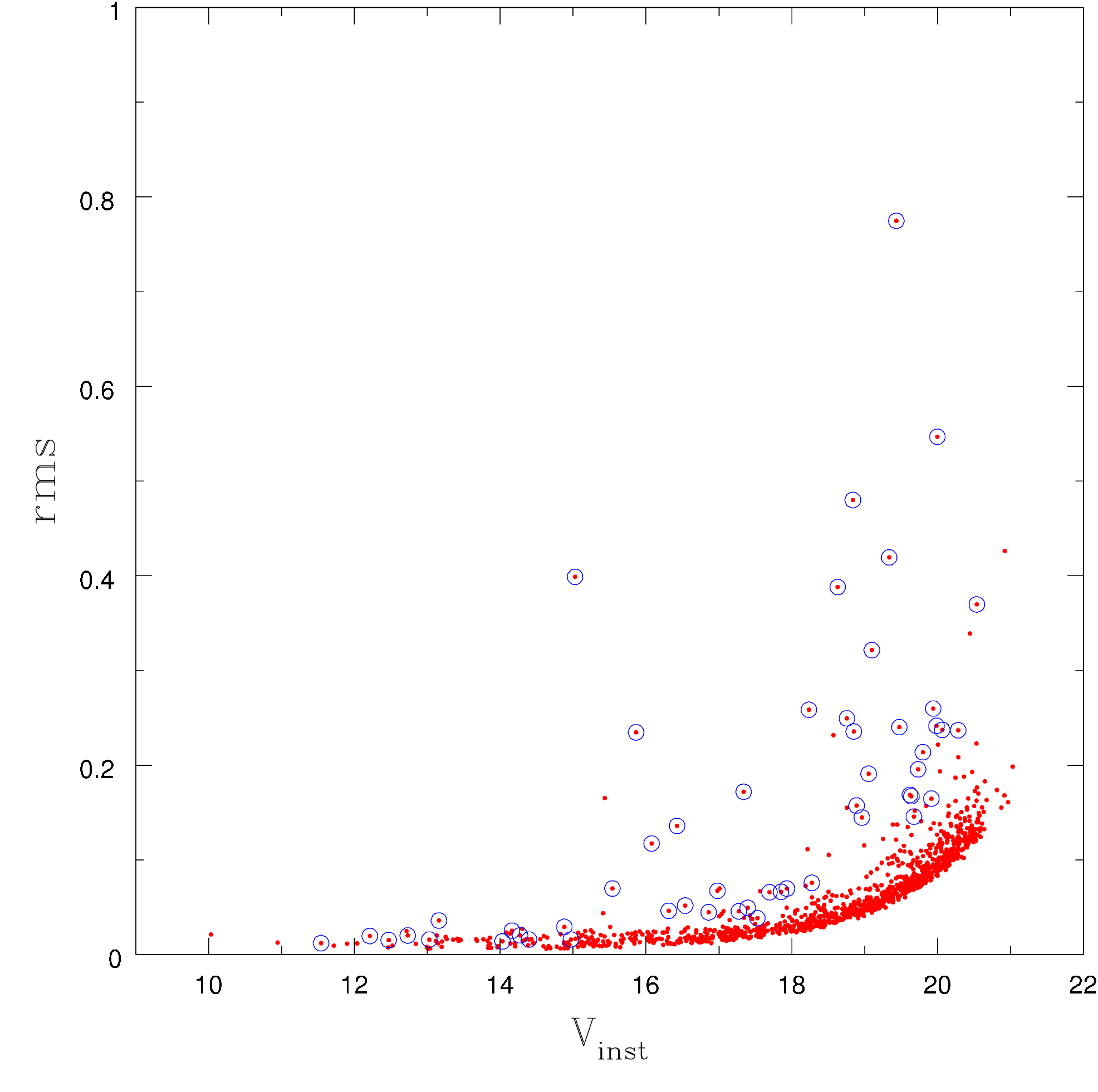}
\caption{The rms for each star as a function of brightness. The open circles represent candidate variables identified in the present work. 
} 
\end{figure}
\begin{figure}
\includegraphics[width=8cm]{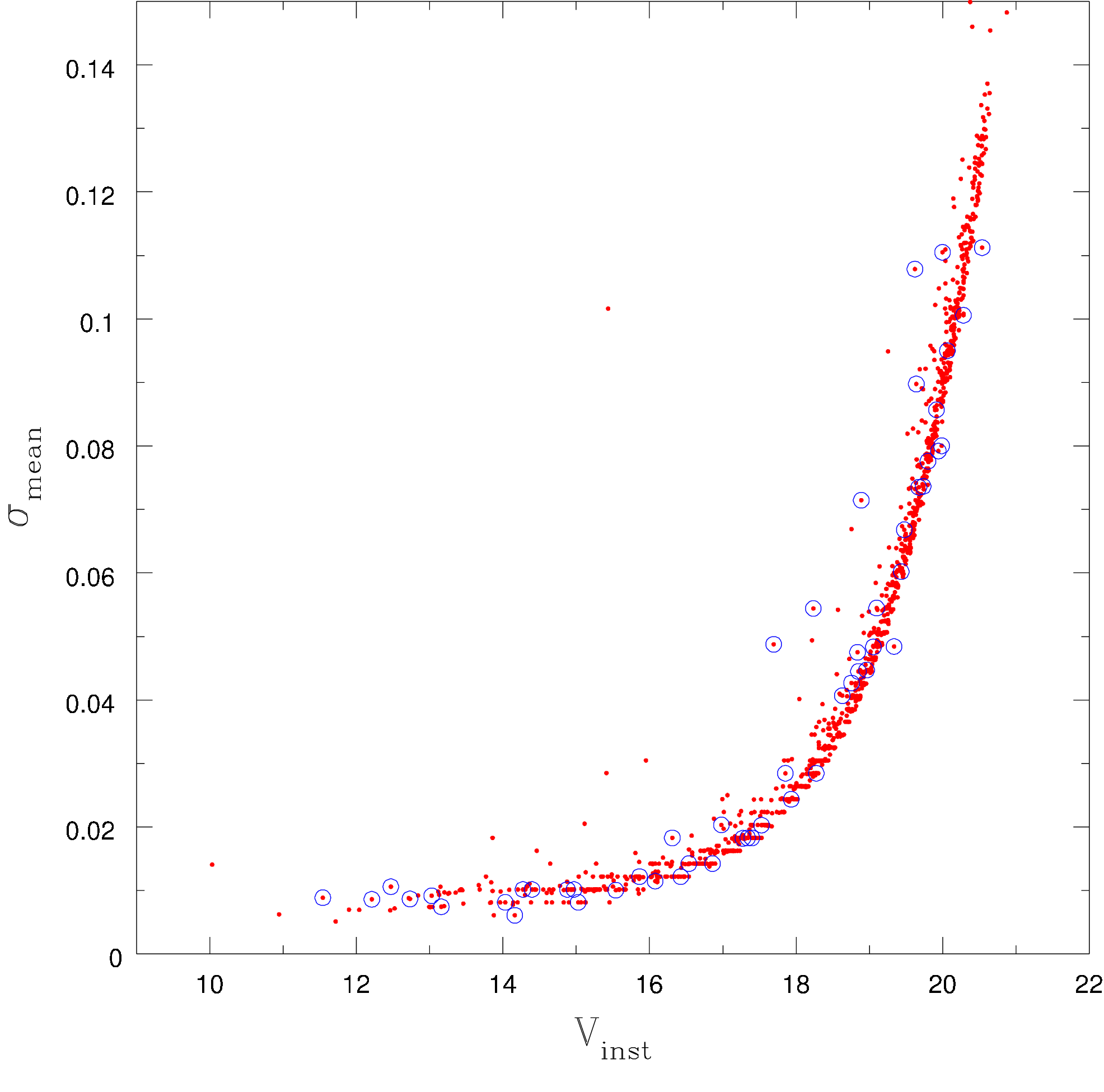}
\caption{Mean photometric errors as a function of brightness. 
Open circles represent
the same as in Fig. 3.}
\end{figure}

\begin{figure}
{
\includegraphics[width=8.cm]{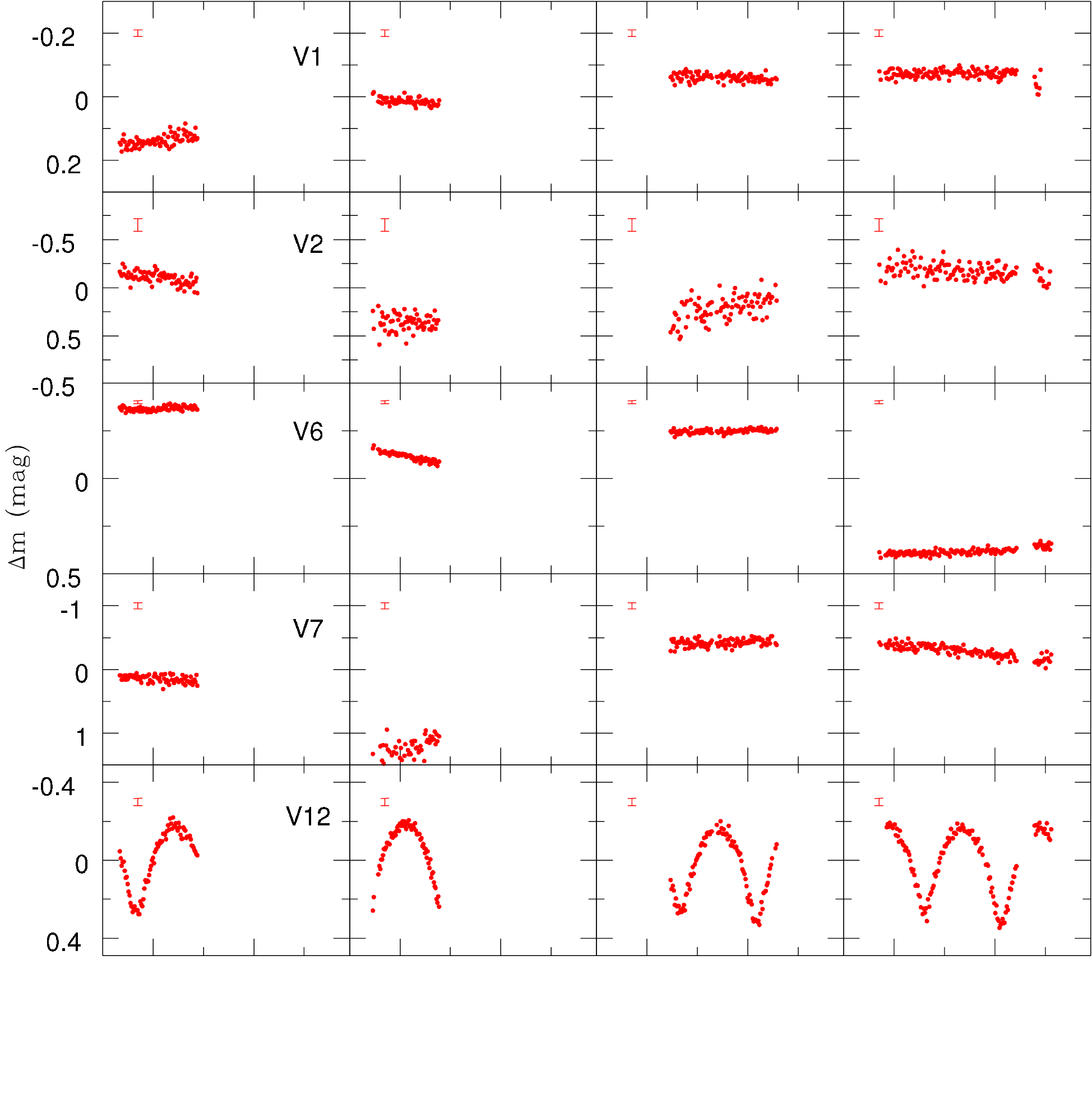}
\includegraphics[width=8.cm]{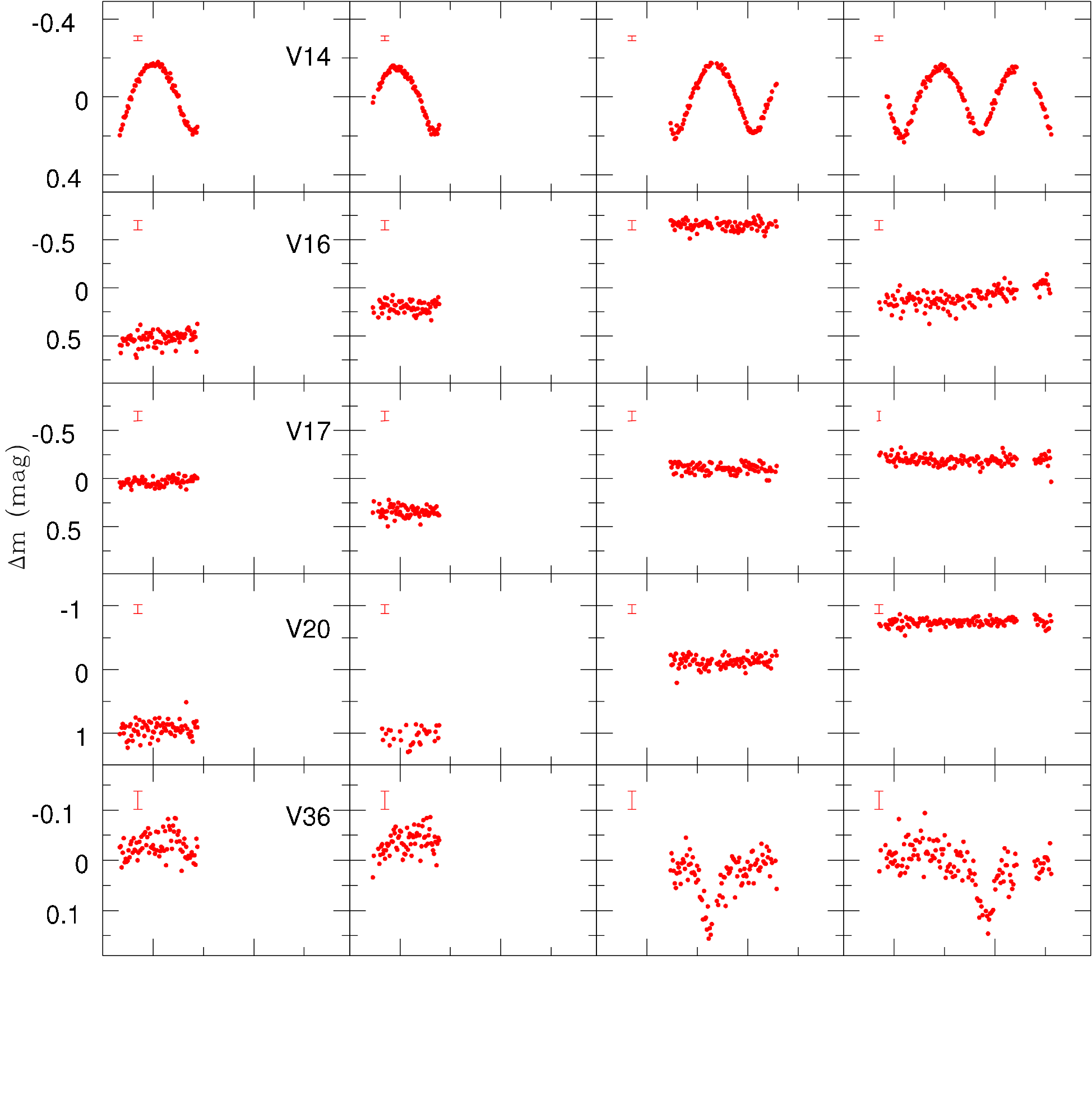}
\includegraphics[width=8.cm]{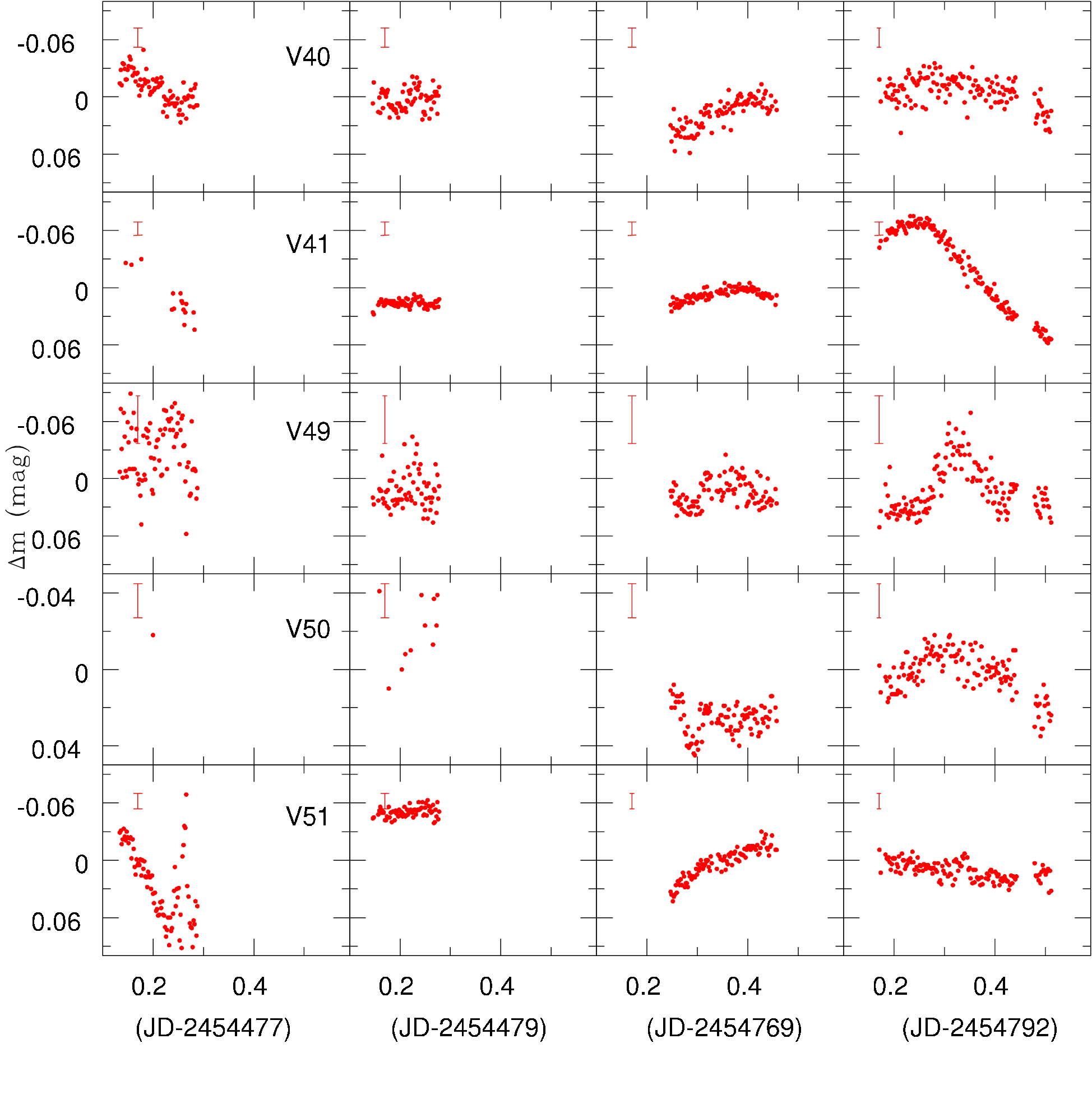}
}
\caption{The light curves of few variable stars identified in the
present work. The typical errors of observations are also shown. The $\Delta$ m represents the differential magnitude in the sense that variable minus comparison star.}
\end{figure}

\subsection{Comparison with Previous Photometry}
We have compared present CCD photometric data with the latest CCD photometry given by Sharma et al. (2007), and its was found that 957 stars are common between these two data sets.
Fig. 2 shows the difference $\Delta$ (present data -literature data) as a function of $V$ magnitude.
 The comparison indicates that the present photometric data are in very good agreement with the CCD photometry by Sharma et al. (2007). The present $V$ magnitudes and $(V-I)$ colours are also in fair agreement with those given  by Sharma et al. (2007).

\subsection{Identification of Variables}
The  mean value of the magnitude and rms of data for each star were estimated using all the 
observations. 
The rms dispersion as a function of instrumental magnitude is shown in Fig. 3, which
indicates that the majority of the stars follow an expected trend i.e., the S/N ratio decreases as stars become fainter.
However, a few stars do not follow the normal trend and exhibit relatively large scatter. These could be either due
to the large photometric errors or due to the variable nature of the stars. We considered a star as variable if its rms is
greater than 3 times of the mean rms of that magnitude bin.
Around 60 candidate variables were thus identified on the basis of the above
mentioned criterion.
A careful inspection of light curves and the location of the sources on the CCD frames rejected a few stars as they were lying near the edge of the CCD. 
Finally, the present sample of variable stars has 53 variable candidates.
The identification number, coordinates and photometric data for these variable stars
are given in Table 2. 
The mean value of photometric error of the data was estimated using the observations of each star. The mean error ($\sigma_{mean}$) as a function of instrumental magnitude is shown in
Fig. 4. The mean error is found to be $\sim$ 0.01 mag at
 magnitude $\le$ 16 mag, whereas its value increases to $\sim$0.1 mag at $\sim$20 mag.
The light curves of a few variables are shown in Fig. 5.
Table 2 also lists NIR and MIR data taken from Prisinzano et al. (2011). The [3.4 $\micron$], [4.6 $\micron$], [12 $\micron$], and [22 $\micron$] data have been taken from the WISE database (http://irsa.ipac.caltech.edu/). 

\begin{figure}
\includegraphics[width=8cm]{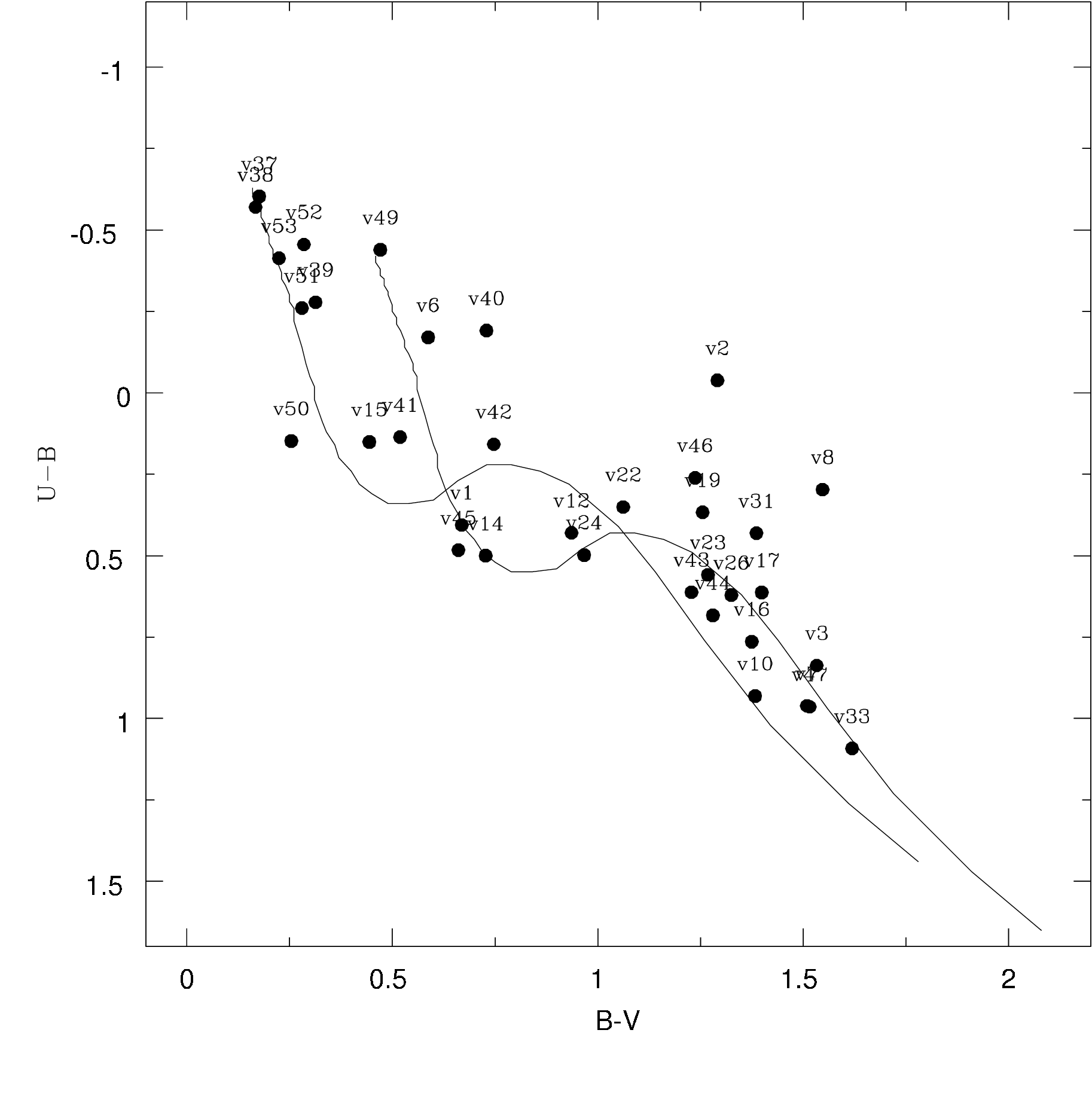}
\caption{$U-B/B-V$ two colour diagram. The UBV data have been taken from 
Sharma et al. (2007). The solid line represents the ZAMS by Giradi et al. (2002) shifted along the reddening vector of 0.72 for $E(B-V)_{min}$= 0.4 mag and $E(B-V)_{max}$= 0.7 mag.}
\end{figure}

\begin{figure}
\includegraphics[width=8cm]{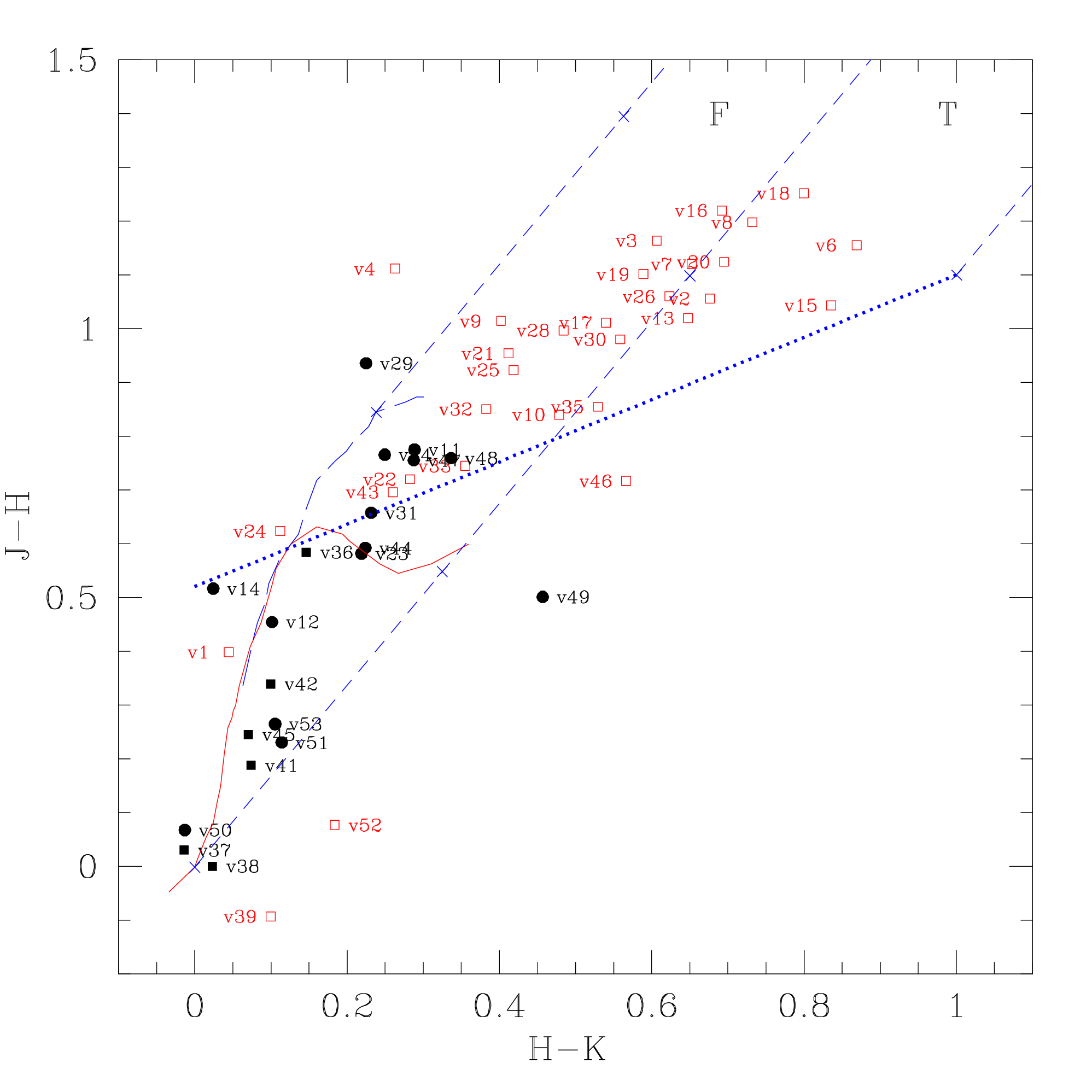}
\caption{($J-H/H-K$) TCD for stars lying in the field of
13$\times$13 arcmin of NGC 1893. $JHK$ data have been taken from Prisinzano et al. (2011). The sequences for dwarfs (solid) and giants (long dashed) are from Bessell \& Brett (1988). The dotted line represents the locus of TTSs (Meyer et al. 1997).
The small dashed lines represent the reddening vectors (Cohen et al. 1981). 
 The crosses on the reddening vectors represent an increment of visual extinction of $A_{V}$ = 5 mag.} 
The open squares and filled circles represent the Class II and Class III sources identified by Prisinzano et al. (2011).
The filled squares represent those stars which were not cross identified in  Prisinzano catalogue.
\end{figure}

\begin{figure}
\includegraphics[width=8cm]{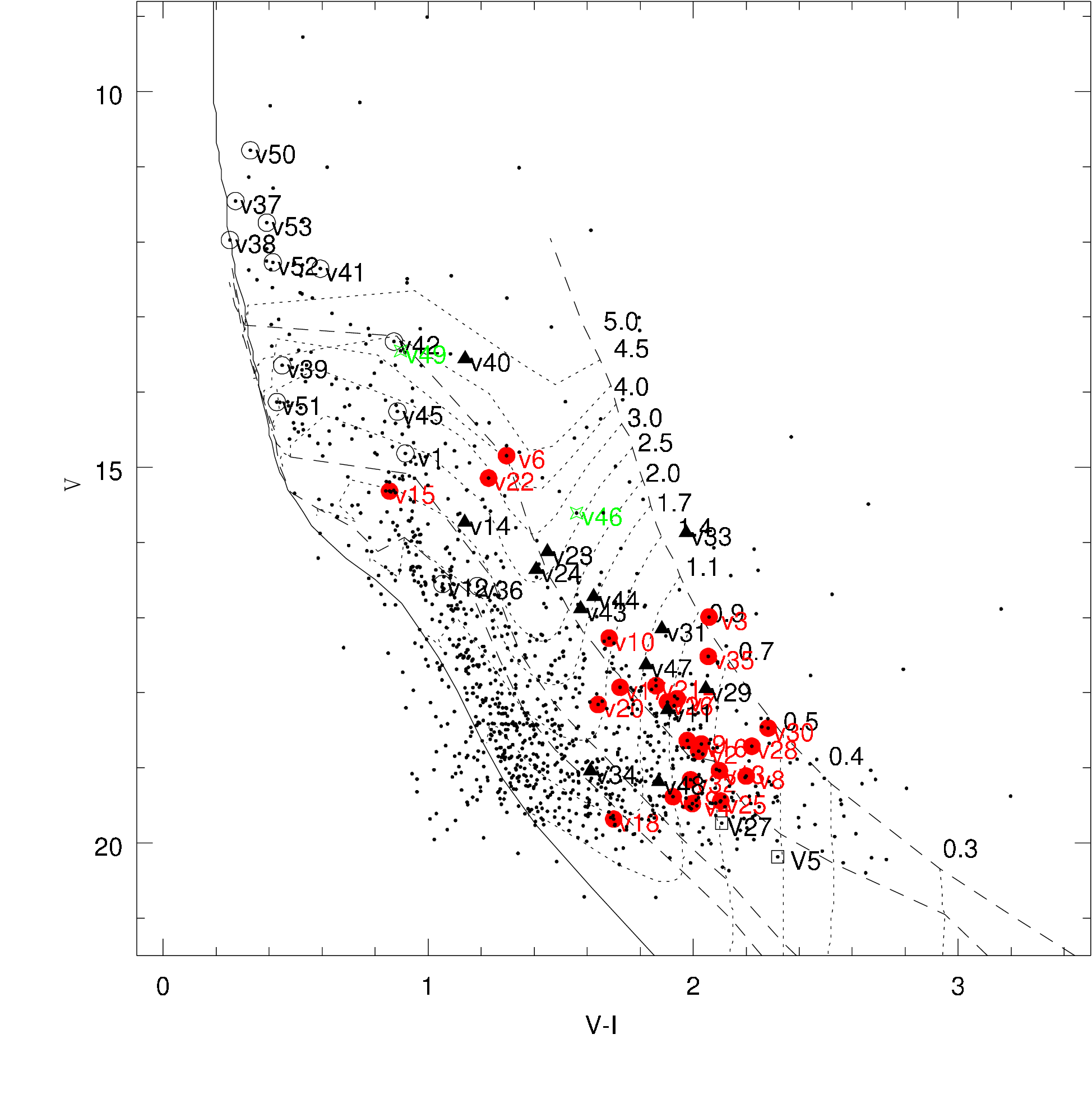}
\caption{$V/V-I$ colour-magnitude diagram for the cluster NGC 1893. 
The small dots represent all the sources in the observed field.
The filled circles and triangles represent CTTSs and WTTSs Table 3. The open circles represent non-PMS stars.
The probable YSOs (V5, V27, V36 and V42) are shown with open squares. Star symbols represent Herbig Ae/Be stars (V46 and V49).
The ZAMS by Girardi et al. (2002)
 and PMS isochrones for
0.1, 1, 5, 10 Myrs by Siess et al. (2000) are shown.
The dotted curves show PMS evolutionary tracks of stars of different masses.
The isochrones and evolutionary tracks  are corrected for the cluster distance and $E(V-I)=0.50$ mag. 
}
\end{figure}

\begin{table*}
\caption{The multiwavelength photometric data of 53 variables in the region of NGC 1893. H$\alpha$ sources are taken from Sharma et al. (2007). The NIR and MIR data have been taken from Prisinzano et al. (2011), when NIR data not available in Prisinzano et al. (2011) the data have been taken from 2MASS point source catalogue (Cutri et al. 2003). The same are marked (**). The [3.4$\micron$], [4.6$\micron$], [12$\micron$], and [22$\micron$] data have been taken from WISE database (http://irsa.ipac.caltech.edu/).  \label{tab:basPar}}

\tiny
\begin{tabular}{llccclclllllp{0.5cm}p{0.5cm}p{0.4cm}p{0.1cm}}
\hline
ID&
$\alpha_{2000}$&
$\delta_{2000}$&
$V$&
$I$&
$J$&
$H$&
$K$&
[3.6$\micron$]&
[4.5$\micron$]&
[5.8$\micron$]&
[8.0$\micron$]&
[3.4$\micron$] &
[4.6$\micron$] &
[12$\micron$] &
[22$\micron$]
\\
&
degree&
degree&
mag&
mag&
mag&
mag&
mag&
mag&
mag&
mag&
mag&
mag&
mag&
mag&
mag 
\\
\hline
  V1&   80.79911 & 33.47988 &               14.845 & 13.901 &  13.299 & 12.918 & 12.869 &   12.303&  11.954&  10.438 &-     &11.350& 11.030& 6.400& 5.297\\                     
  V2$^{\star}$&   80.79100 & 33.48630 &     18.808 & 16.757 &  17.335 & 16.464 & 15.672 &   -     &   -    & -       &-     &-&-&-&-\\                
  V3$^{\star}$&   80.78678 & 33.47591 &     17.022 & 14.932 &  13.638 & 12.653 & 11.943 &   -     &     -  &   7.688 &-     &-&-&-&-\\               
  V4&   80.77605 & 33.46177 &               19.506 & 17.479 &  16.176 & 15.173 & 14.867 &   13.815&  13.430 &  12.751 &12.076&-&-&-&-\\              
  V5&   80.77192 & 33.47964 &               20.214 & 17.865 &   -     &   -    & -      &   -     &  -     &  -      &-     &-&-&-&-\\              
  V6$^{\dagger \star}$&80.76733 & 33.48002& 14.875 & 12.190 &  11.263 & 10.244 & 10.249 &    9.082&   8.402&   8.081 & 7.393&9.125&  8.398&  6.423& 3.653\\     
  V7&   80.76716 & 33.48330 &               18.110 & 16.140 & 14.251  &  13.315 & 12.551&   11.654&  11.186&  10.747&  9.953&\\            
  V8&   80.76635 & 33.49719 &               19.139 & 16.909 & 15.031  &  14.038 & 13.181&   11.941&  11.497&  11.156& 10.425&12.135& 11.513& 9.980&  6.514\\             
  V9&   80.76333 & 33.49077 &               18.666 & 16.658 & 14.955  &  14.069 & 13.600&   12.682&  12.450&  12.259& 11.412&-&-&-&-\\             
  V10&  80.76288 & 33.45175 &               17.303 & 15.590 & 14.155  &  13.446 & 12.887&   12.142&  11.837&  11.572& 10.981&12.540& 12.143& 10.314& 6.870 \\             
  V11&  80.75592 & 33.45502 &               18.254 & 16.321 & 14.825  &  14.141 & 13.804&   13.687&  13.684&  13.561& 13.670&13.255& 13.160& 11.813& 7.436\\             
  V12&  80.75541 & 33.43038 &               16.582 & 15.498 & 14.647  &  14.225 & 14.109&   14.294&  14.202&  14.417&  -    &14.162& 14.058& 12.082& 7.837\\              
  V13&  80.75316 & 33.47177 &               19.067 & 16.938 & 15.187  &  14.344 & 13.585&   12.512&  11.939&  11.393& 10.514&12.630& 11.973& 9.703&  6.380\\              
  V14$^{\star}$&  80.74783 & 33.45941 &     15.761 & 14.593 & 13.930  &  13.436 & 13.410&   13.145&  13.077&  13.160& 13.185&13.196& 13.277& 12.015& 7.632\\   
  V15$^{\star}$&  80.74155 & 33.51174 &     15.346 & 14.462 &  13.411 & 12.581 & 11.602 &   10.580&   9.943&   9.243&  7.824&10.796& 9.867&  6.679&  4.768\\    
  V16$^{\star}$&  80.73480 & 33.44699 &     18.713 & 16.652 &  14.825 & 13.803 & 12.993 &   12.276&  11.847&  11.537& 10.611&12.176& 11.707& 9.455&  7.455\\    
  V17&  80.72919 & 33.46530 &               17.958 & 16.204 & 14.800 &  13.944 & 13.312 &   12.609&  12.180&  11.740& 10.971&12.825& 12.110& 9.820&  8.033\\              
  V18&  80.72777 & 33.45550 &               19.951 & 18.892 & 16.284 &  15.253 & 14.317 &   13.296&  12.842&  12.417& 11.718&13.617& 12.888& 10.878& 8.173\\              
  V19&  80.72396 & 33.42850 &               19.413 & 17.459 & 15.574 &  14.643 & 13.954 &   13.024&  12.658&  12.529& 11.587&12.889& 12.341& 10.889& 7.539\\             
  V20&  80.72127 & 33.48699 &               18.187 & 16.516 & 15.538 &  14.606 & 13.793 &   12.998&  12.697&  12.395& 11.524&13.011& 12.543& 10.279& 7.440\\              
  V21&  80.71241 & 33.42880 &               17.937 & 16.047 & 14.795 &  13.967 & 13.486 &   12.966&  11.685&  12.505& 11.712&-&-&-&-\\              
  V22$^{\star}$&  80.71233 & 33.43002 &     15.173 & 13.915 &  12.990 & 12.356 & 12.027 &   11.218&  10.815& 10.071 &  8.772&10.783& 10.302& 7.662&  5.656\\    
  V23&  80.70471 & 33.43033 &               16.151 & 14.671 &  13.602 & 13.084 & 12.830 &   12.535&  12.716& 12.591 & 12.646&-&-&-&-\\              
  V24&  80.69869 & 33.47938 &               16.387 & 14.950 &  14.768 & 14.191 & 14.062 &   13.082&  13.147& 13.439 &  -    &-&-&-&-\\               
  V25&  80.69780 & 33.45627 &               19.469 & 17.335 &  15.503 & 14.705 & 14.217 &   13.552&  13.205&  12.906& 11.942&13.555& 13.096& 10.911& 7.827\\              
  V26&  80.69466 & 33.49130 &               18.154 & 16.221 &  14.918 & 14.033 & 13.303 &   12.381&  12.016&  11.728& 11.157&12.619& 12.098& 11.693& 8.054\\               
  V27&  80.69864 & 33.40333 &               19.765 & 17.627 &   -     &  -     & -      &   -     &  -     &  -     & -     &-&-&-&-\\               
  V28&  80.69733 & 33.42235 &               18.742 & 16.491 &  15.827 & 14.973 & 14.407 &   13.303&  13.632&  13.502& 12.578&-&-&-&-\\               
  V29&  80.69589 & 33.41677 &               17.976 & 15.898 &  15.142 & 14.297 & 14.035  &  13.596&  13.658&  13.554& -     &-&-&-&-\\               
  V30&  80.66152 & 33.36833 &               18.501 & 16.188 &  14.590 & 13.766 & 13.113 &   12.195&  11.751&  11.360& 10.635&12.110& 11.548& 9.035&  4.718\\              
  V31&  80.66008 & 33.37094 &               17.177 & 15.265 &  13.994 & 13.408 & 13.139 &   13.011&  12.963&  12.857& 12.955&-&-&-&-\\               
  V32$^{**}$&  80.64411 & 33.31941 &               19.188 & 17.168 &  15.484 & 14.662 & 14.258 &   13.400&  13.104&  12.829& 12.003&13.358& 12.930& 8.527&  3.199\\              
  V33&  80.58916 & 33.48075 &               15.900 & 13.896 &  12.263 & 11.620 & 11.206 &   11.027&  11.034&  10.997& 10.931&11.073& 11.107& 9.370&  4.145\\                
  V34&  80.70883 & 33.44688 &               19.068 & 17.425 &  16.270 & 15.587 & 15.297 &   -     &  -     &  -     & -     &-&-&-&-\\               
  V35&  80.69152 & 33.42441 &               17.547 & 15.460 &  14.042 & 13.329 & 12.710 &   11.924&  11.599&  11.130& 10.241&-&-&-&-\\                
  V36$^{**}$&  80.78624 & 33.34811 &               16.611 & 15.395 &  14.511 & 13.946 & 13.783 &   -     &  -     &  -     & -     &13.747& 13.825& 11.207& 6.019\\               
  V37$^{**}$&  80.71744 & 33.38460 &               11.486 & 11.183 &  11.017 &10.986     & 10.987      &   -     &  -     &  -     & -     &10.966& 10.995& 10.364& 5.796\\               
  V38$^{**}$$^{\dagger}$&  80.72360 & 33.39241&    12.005 & 11.723 &   11.575     &  11.573     & 11.536      &   -     &  -     &  -     & -     &11.499& 11.492& 10.613& 5.701\\     
  V39&  80.68433 & 33.41225 &               13.672 & 13.193 &  12.889 & 12.976 & 12.862 &   12.522&  12.497&  12.355& 12.594&12.501& 12.551& 11.251& 7.713\\               
  V40$^{\dagger \star}$&80.67894 & 33.41830&13.585 & 12.416 &  11.419 &  -     & 10.741 &   10.234&   9.970&   9.738& -     &10.373& 9.989&  8.991&  7.554\\  
  V41$^{**}$$^{\dagger}$&  80.66172 & 33.43210 &   12.386 & 11.763 &  11.411     &  11.228     & 11.139      &   -     &  -     &  -     & -     &11.104& 11.131& 11.685& 8.569\\ 
  V42$^{**}$&  80.59608 & 33.43260 &               13.356 & 12.455 &   11.918     &  11.589     & 11.474      &   -     &  -     &  -     & -     &11.390& 11.419& 10.168& 5.182\\              
  V43&  80.76327 & 33.46513 &               16.910 & 15.304 &  14.108 & 13.492 & 13.189 &   13.071&  12.985&  12.045& -     &12.713& 12.638& 8.875&  5.852\\               
  V44&  80.75733 & 33.47427 &               16.752 & 15.097 &  13.804 & 13.277 & 13.018 &   12.694&  12.770&  12.276& 12.257&-&-&-&-\\                 
  V45$^{**}$&  80.60847 & 33.47369 &               14.287 & 13.373 &   12.746     &  12.508     & 12.423     &   -     &  -     &  -     & -     &12.165& 12.207& 10.167& 5.529\\               
  V46&  80.77758 & 33.47830 &               15.639 & 14.049 &  13.044 & 12.466 & 11.804 &   -     &  -     &  -     & -     &-&-&-&-\\   
  V47&  80.70541 & 33.43230 &               17.659 & 15.807 &  14.250 & 13.584 & 13.249 &   13.036&  13.030&  13.043& 13.017&12.942& 12.811&  11.318& 7.723 \\                
  V48&  80.71594 & 33.47783 &               19.214 & 17.314 &  16.191 & 15.530 & 15.138 &   14.815&  14.785&  14.725& -     &15.475& 15.734&  11.516& 8.324\\               
  V49&  80.78121 & 33.47763&                13.480 & 12.554 &  12.244 & 11.847 & 11.313 &   -     &    -   &       -& -     &-&-&-&-\\              
  V50$^{**}$&  80.68846 & 33.37158&                10.811 & 10.452 &  10.129 & 10.062 &  9.989 &    9.840 & 9.914   &  9.903&  9.924&9.862&  9.877&  9.279&   5.257\\                
  V51$^{\dagger}$&  80.71294&  33.42688&    14.161 & 13.702 &  13.520 & 13.309 & 13.178 &   13.148&  13.154&  13.005& -     &-&-&-&-\\  
  V52& 80.68810  & 33.40672 &               12.302 &11.862  &  11.574 & 11.519 & 11.306&    11.331&  11.255& 11.373 & 11.399 &11.241& 11.283& 11.495& 6.439 \\
  V53& 80.57622  & 33.47227 &               11.774 &11.353  &  11.348 & 11.103 & 10.982&    10.957&  10.946& 10.954 &11.029  &10.312& 10.110& 9.693& 5.020 \\
\hline
\end{tabular}

$^{\dagger}$ known variables \\
$^{\star}$ H$\alpha$ star \\
$^{**}$ data from 2MASS catalogue \\
\end{table*}

In Fig. 1, we present the spatial distribution on the DSS image of 
the variable stars detected in the present study.  
The distribution shows a rather aligned distribution from the ionising source of the region (HD 242935) to the nebula Sim129. 
The spatial distribution of variable stars is found to be
similar to that of H$\alpha$ emission line stars  and NIR excess sources
(Marco \& Negueruela 2002; Maheswar et al. 2007; Sharma et al. 2007), which
suggests that these
variable stars should be a part of young stellar population and associated with  the NGC 1893 star forming region.

\section{Membership}
The NIR, MIR TCDs, spectral energy distribution and $(V/V-I)$ CMD have been used to find out
the association of the identified variables with the cluster NGC 1893 and to know the nature of the variables.

\subsection{MIR data}
On the basis of MIR and X-ray data, Prisinzano et al. (2011) have identified 1034 and 442 Class II and Class III young stellar objects (YSO), respectively, in an area of 22.2$\times$18.0 arcmin$^2$.
The classification of identified variables as per Prisinzano et al. (2011) catalogue is given in column 1 of Table 3. It is worthwhile to point out that the identification of Class II and Class III YSOs on the basis of MIR data may be subjected to the contamination due to the field star population (see e.g. Chauhan et al. 2011). Chauhan et al. (2011) have used NIR $J-H/H-K$
 TCD to avoid the field star contamination.

\subsection{Spectral Energy Distribution }
The evolutionary stage of a YSO can also  be studied with the help of the 
spectral energy distribution (SED). 
We adopted the classification scheme of 
Lada et al. (2006), which defines the spectral index $\alpha$=$d\log\lambda(F_{\lambda})/dlog(\lambda)$. 
 We computed the spectral class index $\alpha_{K-8\micron}$, which is the slope of the linear
fit to the fluxes between the K band and IRAC 8$\micron$ band. Objects with $\alpha_{K-8\micron}$ $\ge$ +0.3 are considered as Class I, +0.3 $>$$\alpha_{K-8\micron}$ $\ge$ -0.3 as Flat, -0.3 $>$$\alpha_{K-8\micron}$ $\ge$ -1.8 as Class II and $\alpha_{K-8\micron}$ $<$ -1.8 as Class III sources. 
The above classification needs data from K band to 8 $\micron$ bands, but in the present case some stars 
(namely V1, V12, V24, V29, V35, V40, V48 and V50) have data only up to 5.8 $\micron$. For those stars, $\alpha$ indices were 
obtained using data from  K to the IRAC 5.8$\micron$ band only.  
The $\alpha$ indices obtained from the SED are given in Table 3, which in general are consistent with classification obtained by Prisinzano et al. (2011) on the basis of MIR TCD except for three sources, namely V24, V33 and V39, which were classified as Class 0/I/II sources. 
These sources could be Class III sources. The variables V28, V40 and V43 seem to be border line cases on the basis of the $\alpha$ index. Keeping the error in mind these could be either Class II or Class III sources.  

\begin{figure}
{
\includegraphics[width=8cm,height=8cm]{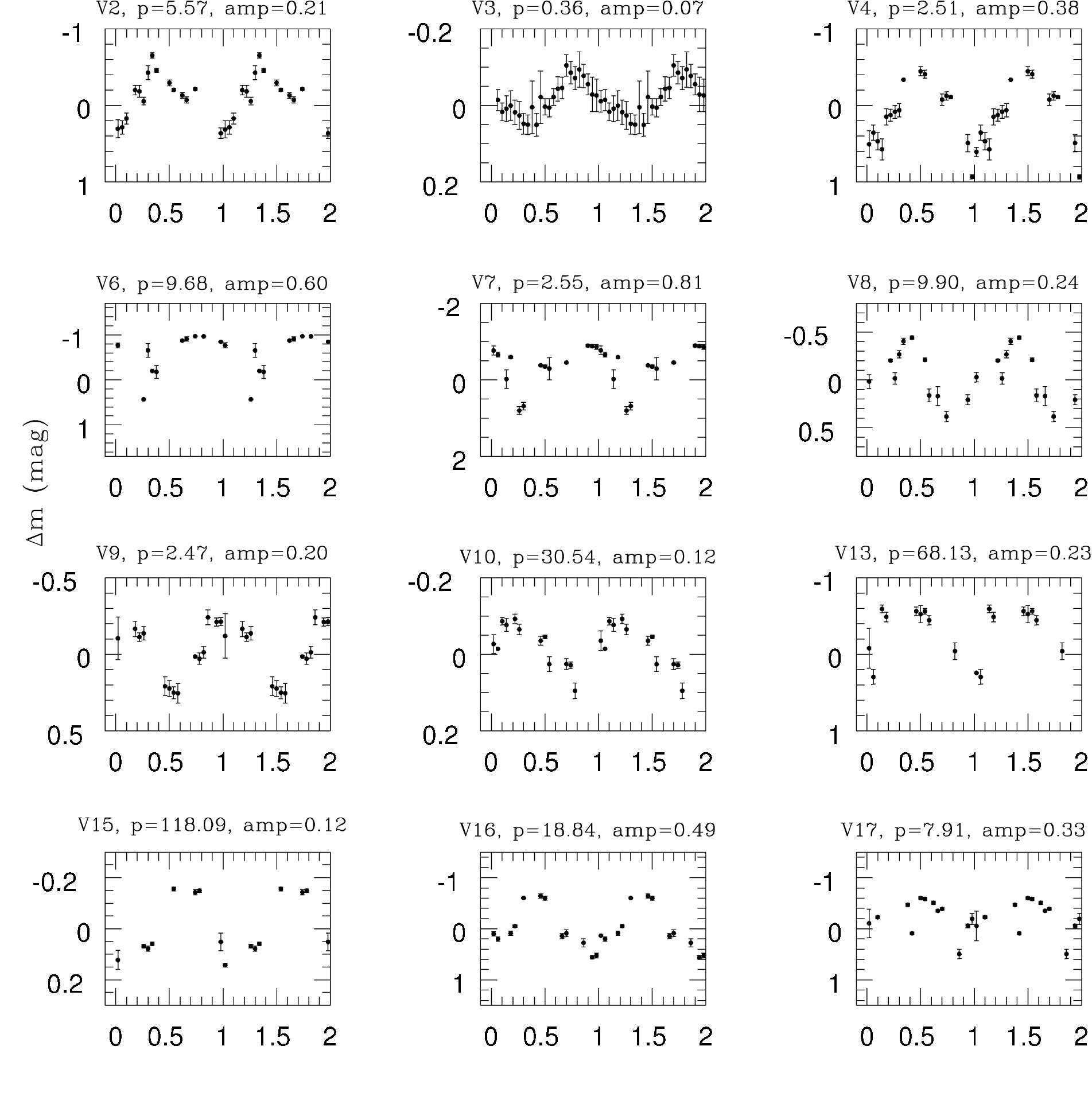}

\includegraphics[width=8cm,height=8cm]{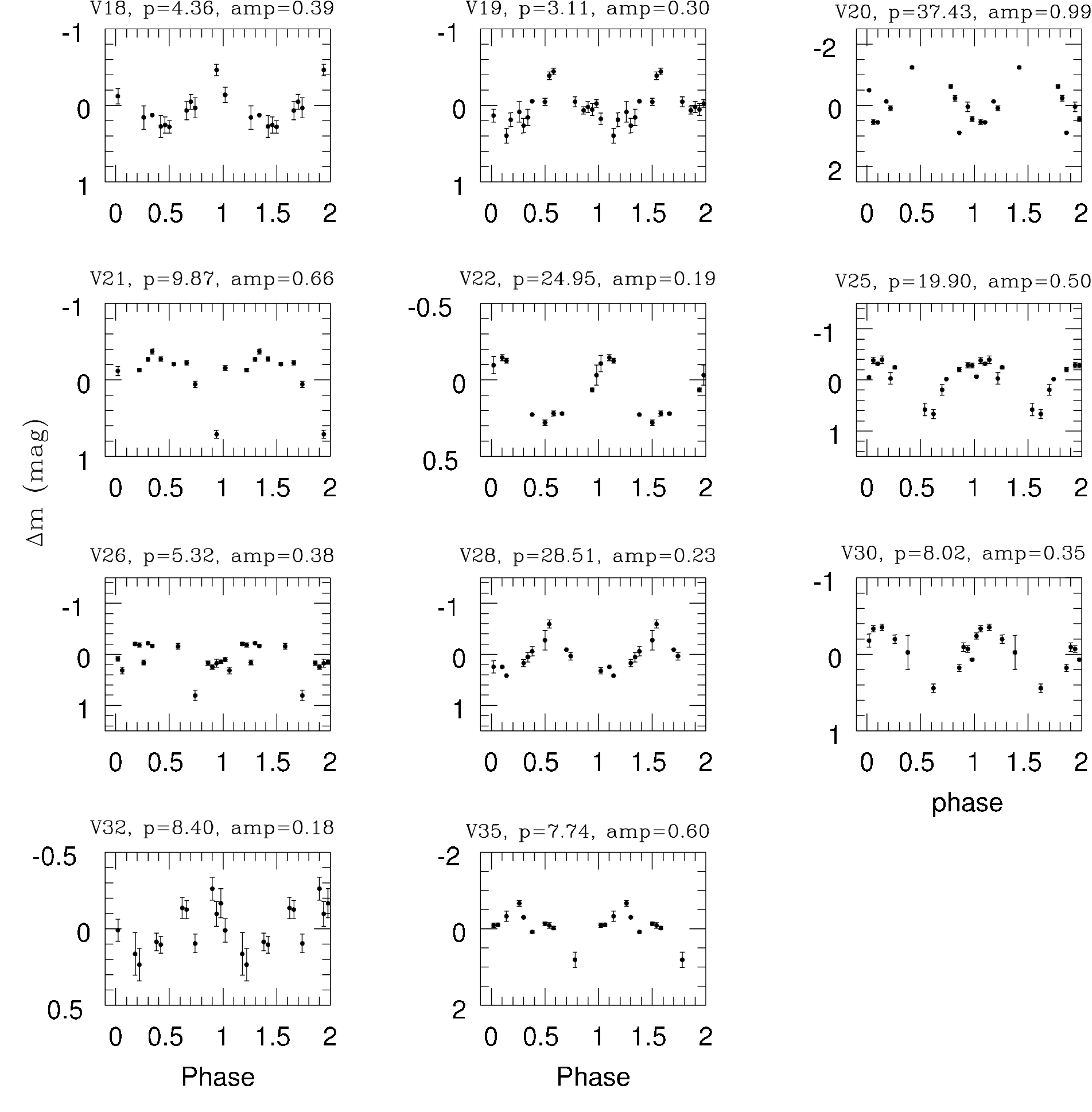}
}
\caption{The phased light curves of probable CTTS candidate variables identified in NGC 1893. The Period is in days and amp is in mag. }
\end{figure}

\begin{figure}
\vbox{
\includegraphics[width=8cm,height=8cm]{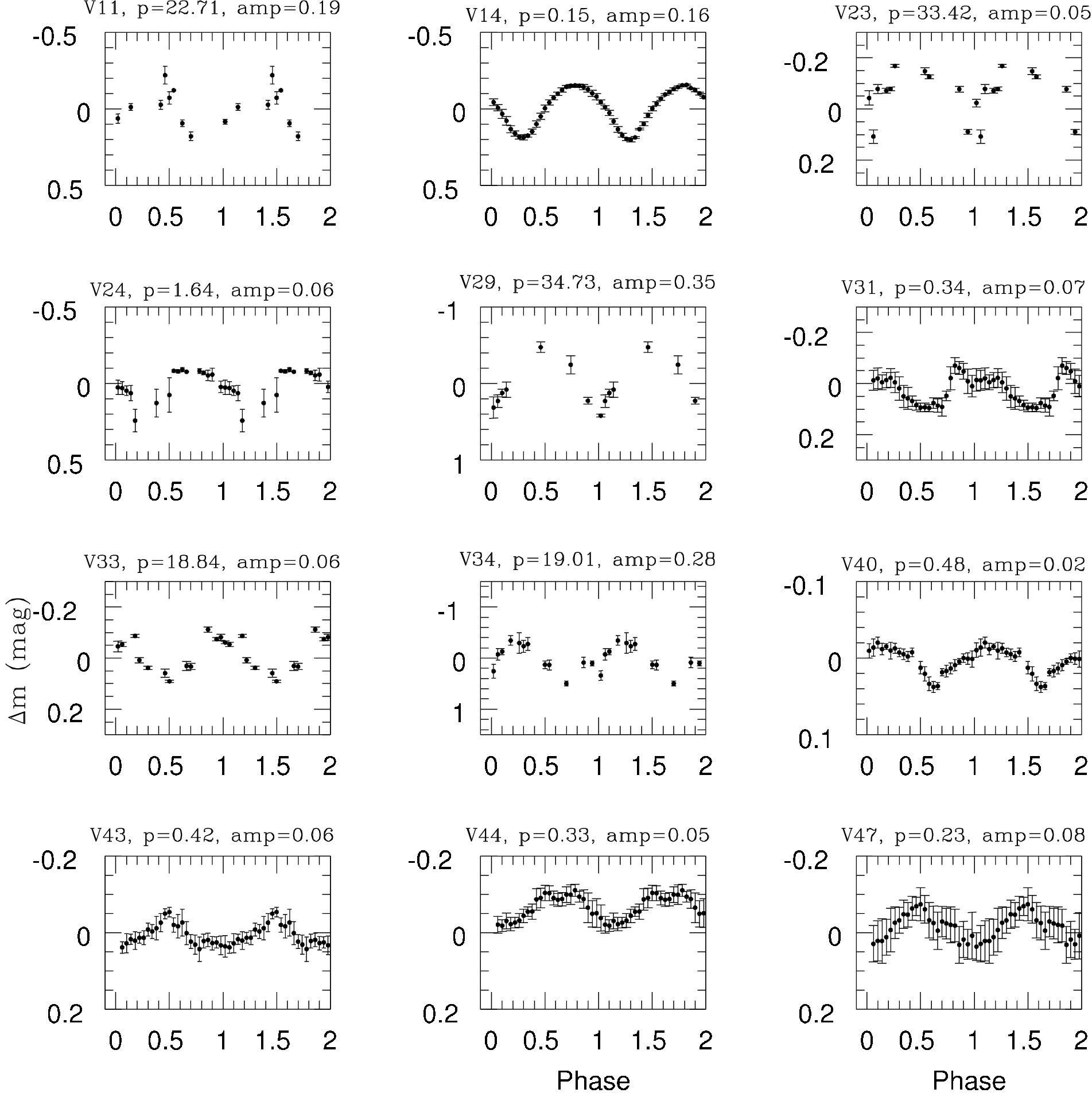}

\hspace{0.35cm}
\includegraphics[width=2.cm,height=2.cm]{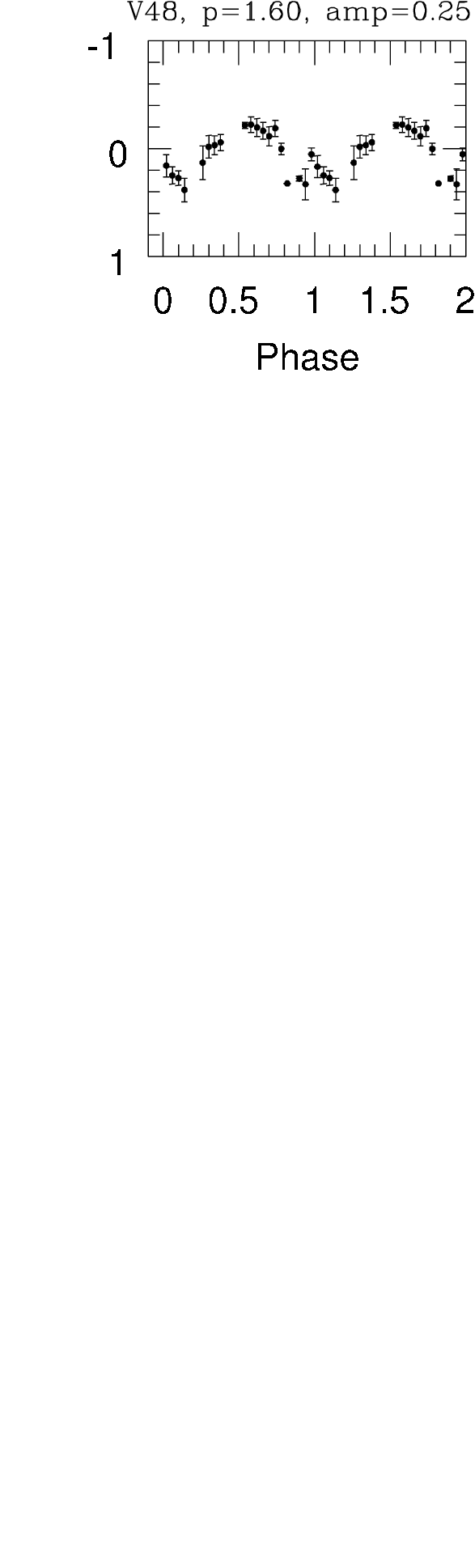}
}
\caption{The phased light curves of probable WTTS candidate variables identified in NGC 1893. The Period is in days and amp is in mag. }
\end{figure}

\begin{figure}
\includegraphics[width=8cm]{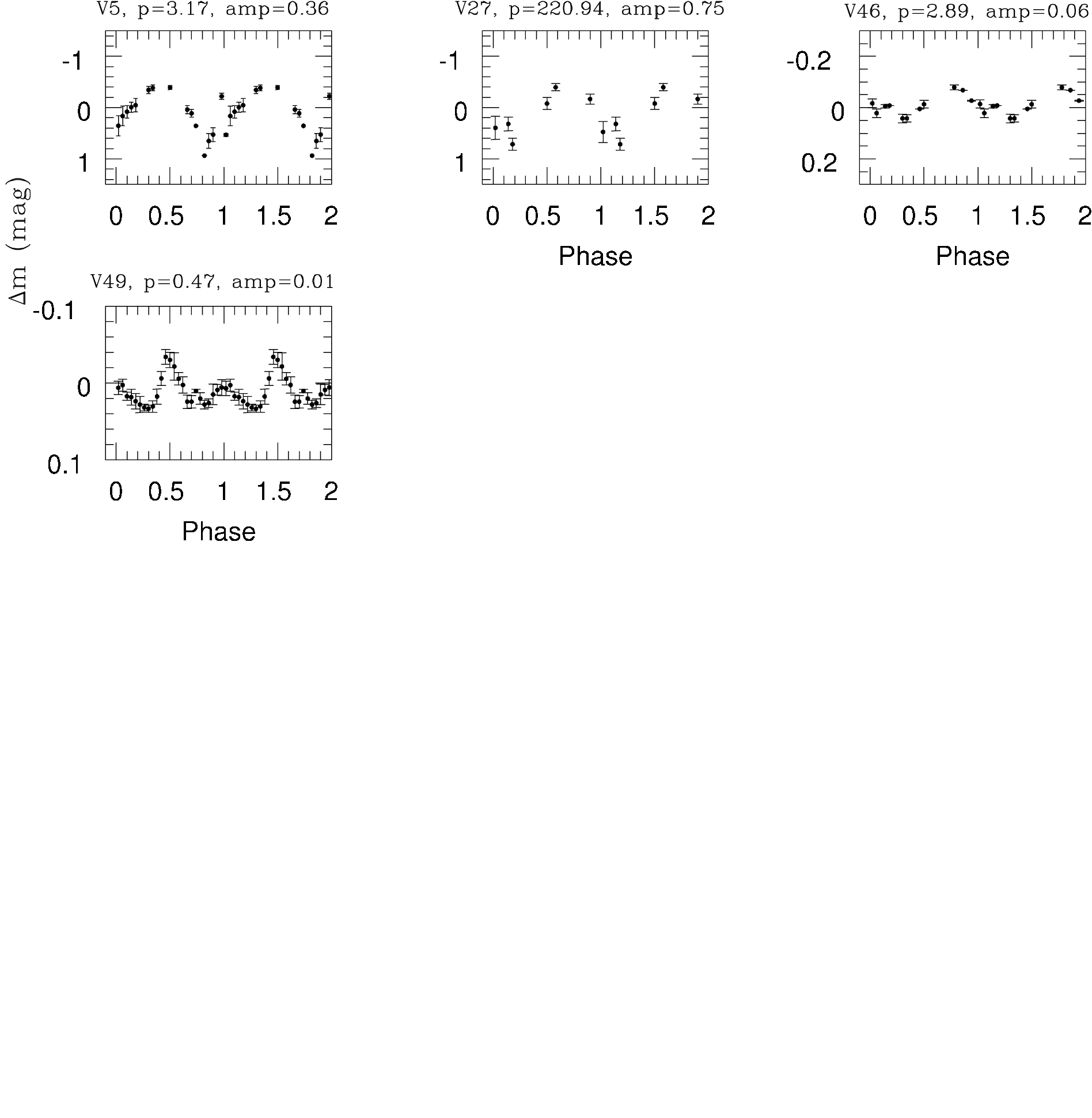}

\caption{The phased light curves of probable YSOs. The Period is in days and amp is in mag.  }
\end{figure}

\begin{figure}

\includegraphics[width=8cm]{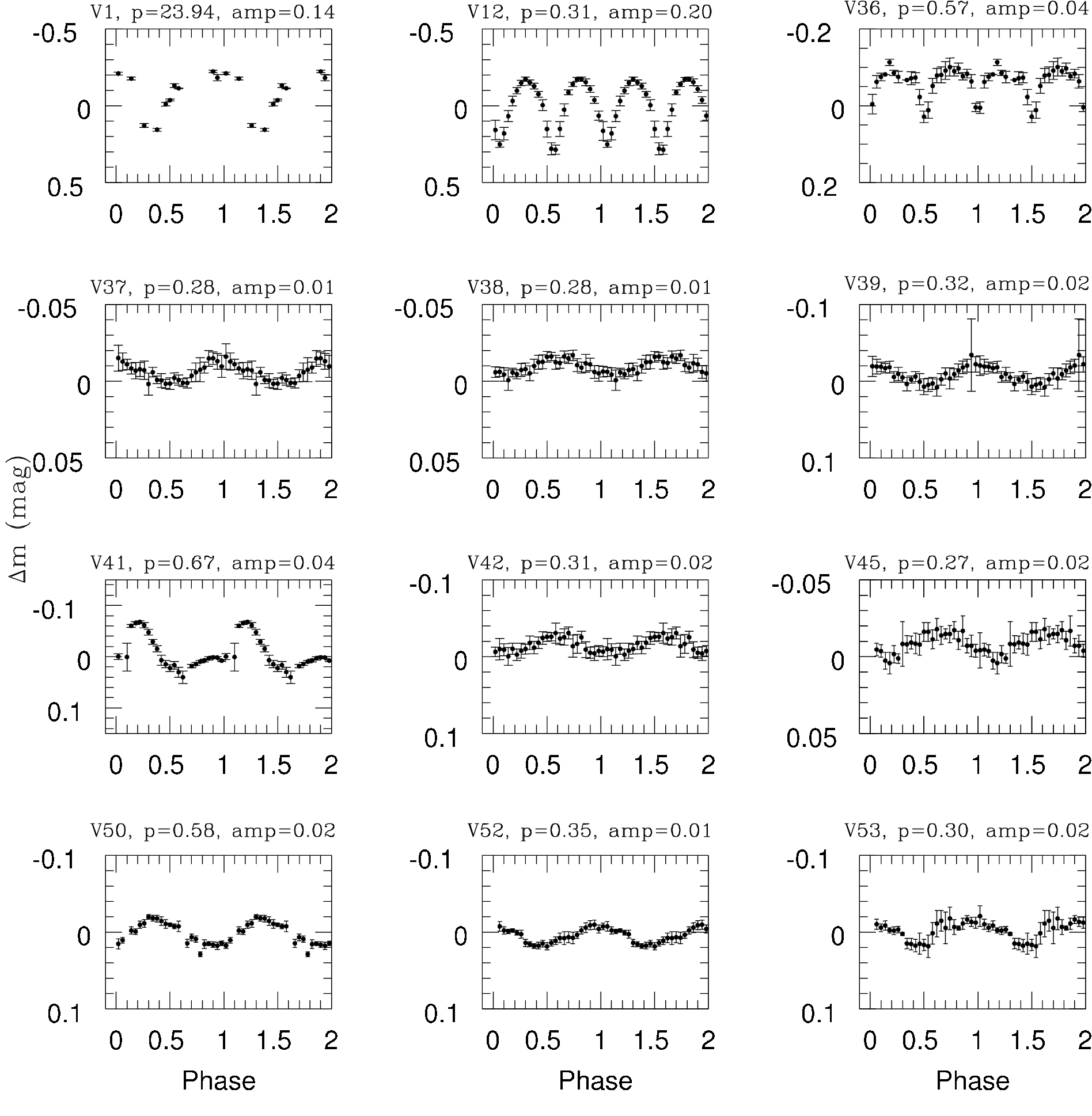}

\caption{The phased light curves of non-PMS stars. The Period is in days and amp is in mag. }
\end{figure}

\subsection{$U-B/B-V$  and $J-H/H-K$ two colour Diagram}
The $U-B/B-V$  TCD for variable candidates
is plotted in Fig. 6.
The $UBV$ data for variable candidates have been taken from 
Sharma et al. (2007). The distribution of stars in the $U-B/B-V$ TCD indicates a variable reddening in the cluster region with $E(B-V)$ from $\sim$0.4 to 0.7 mag. It is not possible to estimate the $E(B-V)$ 
for the YSOs since the $U$ and B band fluxes
may be affected by excess due to accretion which could be an
important origin for
the scattering of the sources in $U-B/B-V$ TCD.

The NIR TCD diagram is also useful to identify the CTTS and
WTTS. Fig. 7 shows the near $J-H/H-K$ TCD. 
The $JHK$ data catalogued by Prisinzano et al. (2011) are in MKO system which 
 were converted to 2MASS system using relations given on the website\footnote[2]{http://www.astro.caltech.edu/~jmc/2mass/v3/transformations/}. After that 
$JHK$ data from both the catalogues (Prisinzano catalogue and 2MASS catalogue) were transformed to CIT system using the relations given on the above mentioned website. The solid
and long dashed lines in Fig. 7 represents unreddened MS and giant loci (Bessell \& Brett 1988) respectively.
The dotted line indicates the intrinsic locus of CTTSs (Meyer et al. 1997).
The parallel dashed lines are the reddening vectors drawn from the tip (spectral type M4) of the giant branch (`left reddening line'), from the base (spectral type A0) of the main-sequence branch (`middle reddening line') and from the tip of the intrinsic CTTS line (`right reddening line'). The extinction ratios $A_J/A_V= 0.265, A_H/A_V= 0.155$ and $A_K/A_V= 0.090$ have been adopted from Cohen et al. (1981).
The sources lying in `F' region could be
either field stars (MS stars, giants) or Class III and Class II sources
with small NIR excesses. The sources lying in the `T' region are considered to be
mostly CTTSs (Class II objects).
Majority of the the Class II sources are distributed in the `T' or near the
middle reddening vector in the `F' region. The locations of V24 and V33 in NIR
 TCD also indicate that these sources should be diskless class III  
instead of Class II as suggested by Prisinzano et al. (2011) on the basis of
MIR TCD. The star V1 is classified as Class I/II source by Prisinzano et al. (2011), whereas its location in NIR TCD as well as in $U-B/B-V$ TCD manifests that it should be late 
 B-type  MS star. 
The location of 
V43 in the NIR TCD suggests that it could be a WTTS. 
The location of V12 in the NIR TCD and $U-B/B-V$ TCD suggests
that it could be an MS star.
Star V39 was classified as Class II by Prisinzano et al. (2011), however its location in NIR TCD and $U-B/B-V$ TCD indicates that it could be an MS or Herbig Be star. 
The location of stars numbered V36, V37, V38, V41, V42, V45 in the NIR TCD as well as in $U-B/B-V$ TCD suggests that these stars could be MS stars.  
The stars
V50 and V51 are found to be diskless/class III source on the basis of Prisinzano et al. (2011) classification as well as on the basis of $\alpha$ index. 
The location of these variables in the NIR TCD (Fig. 7) and in $U-B/B-V$ TCD reveals that these could be reddened O/ B-type stars. The location of V52 and V53 in NIR TCD and $U-B/B-V$ TCD as well as $\alpha$ index indicates that these stars
could be MS stars. Prisinzano et al. (2011) on the basis of
MIR TCD identified V52 and V53 as class II and diskless/class III source respectively.  

The classification of variables adopted in the present study on the basis of MIR data, SEDs, NIR TCD and $U-B/B-V$ TCD is given in the last column of Table 3.

\subsection {$V/V-I$ colour-magnitude diagram}
Fig. 8 shows the $V/V-I$ CMD for all the stars identified (small dots) in the cluster region NGC 1893.
The filled circles and triangles represent probable CTTSs and WTTSs
respectively, whereas open circles represent non-PMS variable stars which are classified in section 3.3 on the basis $J-H/H-K$ and $U-B/B-V$ TCDs. 
The variables V5 and V27 do not have $JHK$ data however they lie in the PMS region of the CMD. These are represented by open squares
and are considered as YSOs. Two stars V46 and V49 which could be Herbig Ae/Be stars are 
represented by star symbols in Fig. 8.
In Fig. 8 we have also plotted theoretical isochrone for 4 Myr for Z=0.02 (continuous line) by Girardi et al. (2002) and PMS isochrones for various ages and evolutionary tracks for various masses by 
Siess et al. (2000). All the isochrones and evolutionary tracks are corrected for the distance (3.25 kpc) and minimum reddening $E(V-I)$=0.50 mag. The minimum value of $E(V-I)$ has been estimated using the relation $E(V-I)/E(B-V)=1.25$ and $E(B-V)=0.40$ mag. 

The ages of YSOs have been derived by comparing their locations with the PMS isochrones, 
whereas masses of PMS variable stars were determined with the help of PMS evolutionary tracks.
The ages of majority of the YSOs are in the range from 0.1 Myr to 5 Myrs which are comparable with the lifetime of TTSs and
 the masses of these sources range from $\sim$0.5 to $\sim$4 M${_\odot}$.
 The associated errors in determination of age and mass can be of two kinds; random errors in observations and systematic errors due to the use of different theoretical evolutionary tracks. We have estimated the effect of random errors in determination of age and mass by propagating the random errors to the observed estimations of $V, V-I$ and $E(V-I)$ by assuming normal error distribution and using the Monte Carlo simulations (see e.g., Chauhan et al. 2009). 
Since we have used model by Siess et al. (2000) only for all the PMS stars, present age and mass estimations are not affected by the systematic errors.
The presence of binaries may be another source of error. The presence of binary will brighten a star, consequently the CMD will yield a lower age estimate. In the case of equal mass binary, we expect an error of $\sim$50 to 60\% in age estimation of the PMS stars. However, it is difficult to estimate the influence of binaries on mean age estimation as the fraction of binaries is not known.
The estimated ages and masses alongwith their errors are given in Table 4.
\begin{table*}
\caption{The classification of variables.}
\tiny
\begin{tabular}{lll}
\hline
Prisinzano et al. (2011) & Spectral index ($\alpha$) &   Present Classification \\
 class0/I/II sources     &                           &                        \\
\hline
V1&                -0.715$\pm$0.713 &           MS \\
V2&                  $-$             &           CTTS \\
V3&                  $-$             &           CTTS \\
V4&                -0.813$\pm$0.132&           CTTS \\
V6&                -0.802$\pm$0.113&           CTTS \\
V7&                -0.963$\pm$0.044&           CTTS \\
V8&                -0.930$\pm$0.096&           CTTS \\
V9&                -1.279$\pm$0.158&           CTTS \\
V10&               -1.471$\pm$0.045&           CTTS \\
V13&               -0.585$\pm$0.030&           CTTS \\
V15&                0.490$\pm$0.155&           CTTS \\
V16&               -0.524$\pm$0.220&           CTTS \\
V17&               -1.221$\pm$0.113&           CTTS \\
V18&               -0.690$\pm$0.159&           CTTS \\
V19&               -1.122$\pm$0.170&           CTTS \\
V20&               -1.109$\pm$0.116&           CTTS \\
V21&               -1.576$\pm$0.476&           CTTS \\
V22&               -0.753$\pm$0.369&           CTTS \\
V24&               -2.128$\pm$0.525&           WTTS \\
V25&               -1.297$\pm$0.173&           CTTS \\
V26&               -1.277$\pm$0.089&           CTTS \\
V28&               -1.722$\pm$0.339&           CTTS \\
V30&               -1.089$\pm$0.039&           CTTS \\
V32&               -1.282$\pm$0.114&           CTTS \\
V33&               -2.608$\pm$0.056&           WTTS \\
V35&               -1.131$\pm$0.140&           CTTS \\
V39&               -2.808$\pm$0.111&           MS/Herbig Be\\
V40&               -1.837$\pm$0.015&           WTTS \\
V43&               -1.862$\pm$0.516&           WTTS  \\
V46&                     $-$         &         Herbig Ae/Be \\
V52&                -2.877$\pm$0.046&           MS \\
\hline
diskless sources&& \\
V11&                -2.629$\pm$0.106&       WTTS \\
V12&                -3.035$\pm$0.082&       MS/Field \\
V14&                -2.741$\pm$0.075&       WTTS \\
V23&                -2.701$\pm$0.130&       WTTS \\
V29&                -2.354$\pm$0.164&       WTTS \\
V31&                -2.575$\pm$0.105&       WTTS \\
V34&                  $-$               &       WTTS \\
V44&                -2.209$\pm$0.137&       WTTS \\
V47&                -2.645$\pm$0.089&       WTTS \\
V48&                -2.389$\pm$0.083&       WTTS \\
V49&                 $-$                &      Herbig Ae/Be \\
V50&                -2.786$\pm$0.066&     MS\\
V51&                -2.678$\pm$0.095&     MS \\
V53&                -2.838$\pm$0.050&      MS/Herbig Be \\
\hline
not classified by Prisinzano et al. (2011)&  & \\
V5&$-$& YSO/Field \\
V27&$-$& YSO/Field \\
V36&$-$& MS/Field\\
V37&$-$& MS \\
V38&$-$& MS\\
V41&$-$& MS/Field\\
V42&$-$& MS\\
V45&$-$& MS\\
\hline
\end{tabular}
\end{table*}

\begin{table}
\caption{The mass, age, amp, and period of the 52 variables.}
\tiny
\begin{tabular}{llllr}
\hline
ID &Mass & Age & Amp. & Period \\
 &$M_{\odot}$ & Myrs & mag & day \\
\hline
V1&-                      &    -               &0.14 & 23.94  \\
V2&0.74$\pm$0.05          &   0.72$\pm$0.03    &0.21 &  5.57  \\
V3& 0.98$\pm$0.01         &   0.10$\pm$0.01    &0.07 &  0.36  \\
V4&0.79$\pm$0.09          &   1.54$\pm$0.44    &0.38 &  2.51  \\
V5&0.51$\pm$0.06          &   1.12$\pm$0.25    &0.36 &  3.17    \\
V6&3.94$\pm$0.15          &   0.64$\pm$0.14    &0.60 &  9.68  \\
V7&0.86$\pm$0.05          &   0.45$\pm$0.03    &0.81 &  2.55 \\
V8&0.58$\pm$0.05          &   0.61$\pm$0.09    &0.24 &  9.90  \\
V9&0.80$\pm$0.05          &   0.68$\pm$0.04    &0.20 &  2.47  \\
V10&1.46$\pm$0.11         &   0.53$\pm$0.09    &0.12 & 30.54 \\
V11&0.90$\pm$0.06         &   0.55$\pm$0.07    &0.19 & 22.71 \\
V12&-                     &   -                &0.20 &  0.31  \\
V13&0.66$\pm$0.05         &   0.75$\pm$0.12    &0.23 & 68.13 \\
V14& 2.48$\pm$0.09        &   3.12$\pm$0.38    &0.16 &  0.15 \\
      &                   &                    &     &  0.17     \\
      &                   &                    &     &  0.15     \\
V15& 2.33$\pm$0.06        &   4.36$\pm$0.55    &0.12 & 118.09\\
V16& 0.73$\pm$0.04        &   0.67$\pm$0.02    &0.49 &  18.84\\
V17& 1.24$\pm$0.08        &   0.79$\pm$0.12    &0.33 &   7.91 \\
V18& 1.14$\pm$0.03        &   8.43$\pm$2.76    &0.39 &   4.36 \\
V19& 0.88$\pm$0.10        &   1.84$\pm$0.55    &0.30 &   3.11 \\
V20& 1.42$\pm$0.09        &   1.49$\pm$0.28    &0.99 &  37.43 \\
V21& 0.99$\pm$0.06        &   0.49$\pm$0.05    &0.66 &   9.87\\
V22& 3.35$\pm$0.19        &   1.22$\pm$0.31    &0.19 &  24.95\\
V23& 2.85$\pm$0.15        &   0.62$\pm$0.11    &0.05 &  33.42\\
V24& 2.75$\pm$0.08        &   0.93$\pm$0.21    &0.06 &  1.64 \\
V25& 0.66$\pm$0.05        &   1.02$\pm$0.21    &0.50 &  19.90\\
V26& 0.91$\pm$0.05        &   0.51$\pm$0.06    &0.38 &   5.32\\
V27& 0.67$\pm$0.07        &   1.40$\pm$0.38    &0.75 & 220.94\\
V28& 0.60$\pm$0.07        &   0.42$\pm$0.24    &0.23 &  28.51\\
V29& 0.77$\pm$0.06        &   0.30$\pm$0.12    &0.35 &  34.73\\
V30& 0.57$\pm$0.01        &   0.11$\pm$0.03    &0.35 &   8.02\\
V31& 1.06$\pm$0.06        &   0.28$\pm$0.02    &0.07 &   0.34\\
V32& 0.78$\pm$0.06        &   1.07$\pm$0.13    &0.18 &   8.40\\
V33& 1.60$\pm$0.01        &   0.10$\pm$0.01    &0.06 &  18.84\\
V34& 1.33$\pm$0.03        &   5.27$\pm$1.35    &0.28 &  19.01\\
V35& 0.81$\pm$0.02        &   0.16$\pm$0.07    &0.60 &   7.74\\
V36& -                    &   -                &0.04 &   0.57\\
V37& -                    &   -                &0.01 &   0.28\\
V38& -                    &   -                &0.01 &   0.28\\
V39& -                    &   -                &0.02 &   0.32\\
V40& 4.83$\pm$0.06        &   0.45$\pm$0.02    &0.02 &   0.48\\
V41& -                    &   -                &0.04 &   0.67\\
V42& -                    &   -                &0.02 &   0.31\\
V43& 1.90$\pm$0.13        &   0.60$\pm$0.10    &0.06 &   0.42 \\
V44& 1.78$\pm$0.12        &   0.42$\pm$0.06    &0.05 &   0.33 \\
V45& -                    &   -                &0.02 &    0.27 \\
V46& 2.74$\pm$0.17        &   0.27$\pm$0.03    &0.06 &    2.89 \\
V47& 1.09$\pm$0.08        &   0.44$\pm$0.05    &0.08 &    0.23 \\
V48& 0.95$\pm$0.09        &   1.73$\pm$0.46    &0.25 &    1.60  \\
V49& 4.61$\pm$0.04        &   0.58$\pm$0.03   &0.01 &    0.47 \\
V50& -                    &   -                &0.02 &    0.58 \\
V52& -                    &   -                &0.01 &   0.35 \\
V53& -                    &   -                &0.02 &   0.30 \\
\hline
\end{tabular}
\end{table}

\begin{table*}
\caption{Physical parameters of stars obtained from SEDs.}
\tiny
\begin{tabular}{lllll}
\hline
ID&         Disk mass &  $\sigma$ & Disk Accretion Rate & $\sigma$ \\
  &         $M_{\odot}$ & $M_{\odot}$ & $M_{\odot}$/yr & $M_{\odot}$/yr \\   
\hline
 V1     &   2.235$\times$10$^{-4}$  &   5.018$\times$10$^{-4}$   & 8.493$\times$10$^{-10}$  &  8.461$\times$10$^{-10}$ \\
 V4     &   4.390$\times$10$^{-3}$  &   1.545$\times$10$^{-2}$   & 2.432$\times$10$^{-08}$  &  2.542$\times$10$^{-08}$ \\
 V6     &   8.818$\times$10$^{-3}$  &   1.878$\times$10$^{-2}$   & 2.395$\times$10$^{-08}$  &  2.326$\times$10$^{-08}$ \\
 V7     &   2.684$\times$10$^{-2}$  &   4.083$\times$10$^{-2}$   & 6.293$\times$10$^{-07}$  &  7.073$\times$10$^{-07}$ \\
 V8     &   1.738$\times$10$^{-2}$  &   2.449$\times$10$^{-2}$   & 1.040$\times$10$^{-07}$  &  1.015$\times$10$^{-07}$ \\
 V9     &   7.637$\times$10$^{-3}$  &   2.108$\times$10$^{-2}$   & 6.864$\times$10$^{-08}$  &  6.735$\times$10$^{-08}$ \\
 V10    &   6.436$\times$10$^{-3}$  &   1.346$\times$10$^{-2}$   & 3.262$\times$10$^{-08}$  &  3.199$\times$10$^{-08}$ \\
 V11    &   1.075$\times$10$^{-2}$  &   1.600$\times$10$^{-2}$   & 1.428$\times$10$^{-08}$  &  1.406$\times$10$^{-08}$ \\
 V13    &   2.288$\times$10$^{-2}$  &   4.271$\times$10$^{-2}$   & 1.807$\times$10$^{-07}$  &  1.772$\times$10$^{-07}$ \\
 V14    &   1.004$\times$10$^{-3}$  &   1.150$\times$10$^{-3}$   & 2.607$\times$10$^{-09}$  &  2.487$\times$10$^{-09}$ \\
 V15    &   6.723$\times$10$^{-3}$  &   1.339$\times$10$^{-2}$   & 2.826$\times$10$^{-09}$  &  2.754$\times$10$^{-09}$ \\
 V16    &   1.807$\times$10$^{-2}$  &   2.208$\times$10$^{-2}$   & 1.469$\times$10$^{-07}$  &  1.411$\times$10$^{-07}$ \\
 V17    &   1.322$\times$10$^{-2}$  &   2.050$\times$10$^{-2}$   & 4.159$\times$10$^{-08}$  &  3.987$\times$10$^{-08}$ \\
 V18    &   1.404$\times$10$^{-2}$  &   1.992$\times$10$^{-2}$   & 2.444$\times$10$^{-08}$  &  2.383$\times$10$^{-08}$ \\
 V19    &   4.461$\times$10$^{-3}$  &   7.915$\times$10$^{-3}$   & 2.188$\times$10$^{-08}$  &  2.129$\times$10$^{-08}$ \\
 V20    &   9.360$\times$10$^{-4}$  &   8.594$\times$10$^{-4}$   & 3.903$\times$10$^{-09}$  &  3.872$\times$10$^{-09}$ \\
 V21    &   5.105$\times$10$^{-3}$  &   1.449$\times$10$^{-2}$   & 2.345$\times$10$^{-08}$  &  2.449$\times$10$^{-08}$ \\
 V22    &   1.866$\times$10$^{-2}$  &   3.846$\times$10$^{-2}$   & 1.438$\times$10$^{-07}$  &  1.422$\times$10$^{-07}$ \\
 V23    &   1.296$\times$10$^{-3}$  &   5.484$\times$10$^{-3}$   & 3.087$\times$10$^{-09}$  &  5.101$\times$10$^{-09}$ \\
 V24    &   2.251$\times$10$^{-3}$  &   9.066$\times$10$^{-3}$   & 3.481$\times$10$^{-09}$  &  3.565$\times$10$^{-09}$ \\
 V25    &   2.830$\times$10$^{-3}$  &   5.891$\times$10$^{-3}$   & 2.855$\times$10$^{-09}$  &  2.802$\times$10$^{-09}$ \\
 V26    &   5.537$\times$10$^{-3}$  &   8.124$\times$10$^{-3}$   & 1.793$\times$10$^{-08}$  &  1.778$\times$10$^{-08}$ \\
 V28    &   3.943$\times$10$^{-3}$  &   1.686$\times$10$^{-2}$   & 2.608$\times$10$^{-08}$  &  2.770$\times$10$^{-08}$ \\
 V29    &   1.660$\times$10$^{-3}$  &   6.710$\times$10$^{-3}$   & 3.390$\times$10$^{-09}$  &  3.985$\times$10$^{-09}$ \\
 V30    &   2.991$\times$10$^{-2}$  &   4.062$\times$10$^{-2}$   & 2.004$\times$10$^{-07}$  &  1.953$\times$10$^{-07}$ \\
 V31    &   1.289$\times$10$^{-3}$  &   5.036$\times$10$^{-3}$   & 3.616$\times$10$^{-09}$  &  3.839$\times$10$^{-09}$ \\
 V32    &   2.737$\times$10$^{-2}$  &   3.848$\times$10$^{-2}$   & 4.517$\times$10$^{-07}$  &  4.457$\times$10$^{-07}$ \\
 V33    &   1.605$\times$10$^{-2}$  &   8.095$\times$10$^{-3}$   & 5.469$\times$10$^{-08}$  &  5.431$\times$10$^{-08}$ \\
 V35    &   3.600$\times$10$^{-2}$  &   2.876$\times$10$^{-2}$   & 1.045$\times$10$^{-07}$  &  9.988$\times$10$^{-08}$ \\
 V40    &   1.283$\times$10$^{-2}$  &   1.382$\times$10$^{-2}$   & 2.203$\times$10$^{-09}$  &  2.159$\times$10$^{-09}$ \\
 V43    &   1.283$\times$10$^{-2}$  &   1.382$\times$10$^{-2}$   & 1.342$\times$10$^{-08}$  &  1.315$\times$10$^{-08}$ \\    
 V44    &   2.916$\times$10$^{-3}$  &   9.517$\times$10$^{-3}$   & 1.201$\times$10$^{-08}$  &  1.652$\times$10$^{-08}$ \\  
 V47    &   6.960$\times$10$^{-4}$  &   1.494$\times$10$^{-3}$   & 1.613$\times$10$^{-09}$  &  1.565$\times$10$^{-09}$ \\ 
 V48    &   4.096$\times$10$^{-3}$  &   7.598$\times$10$^{-3}$   & 3.721$\times$10$^{-08}$  &  3.684$\times$10$^{-08}$ \\  
\hline
\end{tabular}
\end{table*}

\section{Period Determination}
We used the Lomb-Scargle (LS) periodogram (Lomb 1976; Scargle 1982) to determine the most probable
period of a variable star. The LS method is useful to estimate 
periodicities even in the case of unevenly spaced data. We used the algorithm
available at the Starlink\footnote[3]{http://www.starlink.uk} software 
database.
The periods were further verified with the software period04\footnote[4]{http://www.univie.ac.at/tops/Period04} (Lenz \& Breger 2005). 
The software period04 provides the frequency and semi-amplitude of the variability in a light curve.
Periods derived from the LS method and Period04 generally
matched well.
The light curves of variable stars are folded with their estimated period.
The phased light curves of variable stars identified as CTTSs, WTTSs, probable YSOs and non-YSOs are shown
in Figs. 9, 10, 11 and 12, respectively, where
averaged differential magnitude 
in 0.04 phase bin along with $\sigma$ error bars have been plotted.

Zhang et al. (2008) identified 10  B-type variable 
stars towards the direction of NGC 1893. Of these, 7 stars (no 6, 8, 10, 3, 5, 9 and 1 in Zhang et al. 2008) 
are 
 common with our present work (V6, V38, V40, V41, V51, V52 and V53 ).
Three variable stars, namely star nos. 2, 4, 7 of Zhang et al. (2008) were not observed by us.

The star V6 has been classified as a CTTS by Prisinzano et al. (2011) as well as in the present work.
Its light curve suggests that it could be an eclipsing binary. 
Zhang et al. (2008) have reported it an
INSA type variable.
They derived two values for the period as 5.747 day and 2.041 day, whereas the values for amplitudes are 0.029 mag and 0.010 mag. 
Our data suggest a period of 9.68 day and a much larger amplitude, 0.60 mag.

We confirm the variability of V38 suspected by Zhang et al. (2008). Its
location in the $V/V-I$ CMD suggests an
MS star.
Massey et al. 1995 found it to be a  B-type star. 

Zhang et al. (2008) also found photometric variations in V40 and V41.
The present study confirms their variability. The period estimates (0.257, 0.273 and 5.747 day) for V41 by Zhang et al. (2008) do not match with the present period
estimate (0.67 day).

Star V51 seems to be a  B-type MS star.
Marco \& Negueruela (2002) also found it to be a  B-type star.
The variability in star V51 has also been detected by Zhang et al. (2008) and they found a period of
10.0 day. The present work suggests that it could be a irregular type variable (cf. Section 6.1).

 Stars V52 and V53 could be MS  B-type/Herbig Be stars. Zhang et al. (2008) identified star V53 as  B-type variable while in case of V52 they suspected it
to be a variable. They estimated the period and amplitude for V53 as 0.26 day and 0.01 mag. In the present work we confirm the variability of V52.

\begin{figure}
{
\includegraphics[width=8cm]{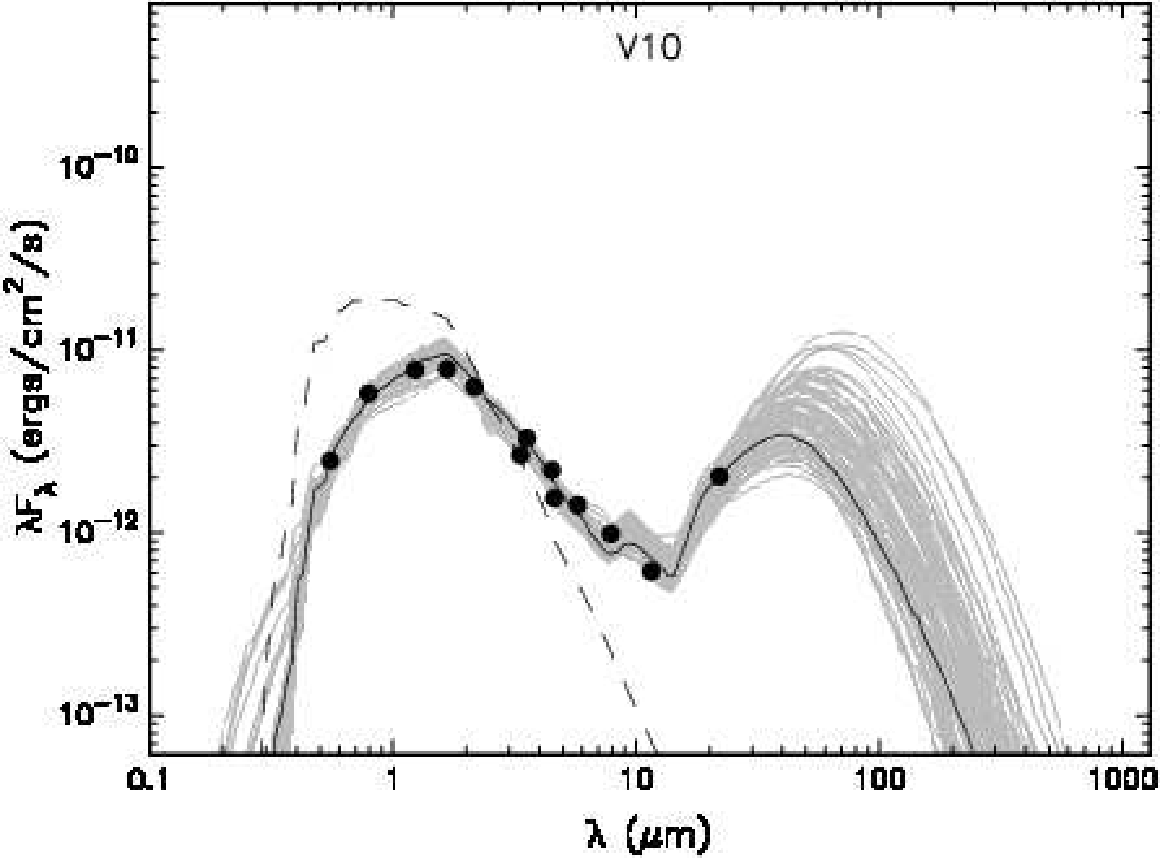}
\includegraphics[width=8cm]{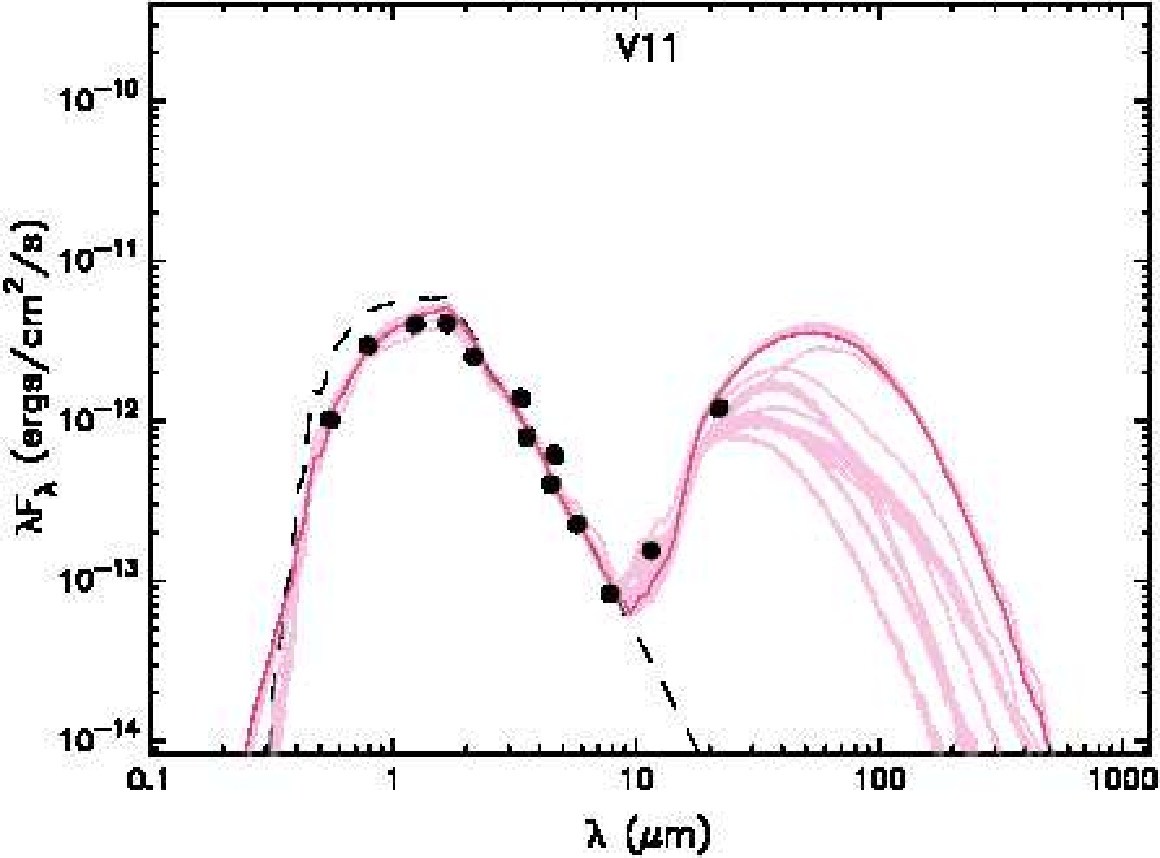}
}
\caption{The spectral energy distribution for V10 and V11 PMS stars. The filled
circles denote the observed flux. The black line shows the best fit
and the grey lines show possible range of good fits. The dashed line shows the stellar photosphere corresponding to the central source of the best-fit model. The results of the best-fit model in each case are given in Table 5.}
\end{figure}
\section {Physical state of variable stars}
In order to characterize the circumstellar disk properties of YSOs
we have analysed the SEDs of the young population associated with the young open cluster NGC 1893.
To construct an SED we used 
multiwavelength i.e optical ($V, I$), NIR ($JHK$), mid IR $Spitzer$ IRAC (at four wavelengths 3.6, 4.5, 5.8 and 8 micron), NASA's Wide-field Infrared Survey Explorer (WISE) at 3.4, 4.6, 12, and 22 micron data and fitting
tool of Robitaille et al. (2007).
Of the 53 detected variables, 45 stars are found in the Prisinzano catalogue.
Only 34 stars of  53 variable candidates have their WISE counterparts.
Two probable YSOs V5 and V27 were not cross identified in WISE catalogue also,
therefore we could not construct the SEDs for these stars.
The SED fitting tool of Robitaille et al. (2007) fits thousands of models to the observed SED
simultaneously. 
This SED fitting tool determines their physical parameters like interstellar extinction, temperature,
disk mass, disk mass accretion rate, etc.
The SED fitting tool needs magnitudes, interstellar extinction $A_{V}$ and distance of the object as input parameters.
The SED fitting tool fits each of the models to the data, allowing both the distance and external foreground extinction to be free parameters. 
 As discussed in Section 3.3 the estimation of $E(B-V)$ for YSOs is not possible on the basis of $U-B/B-V$ TCD,  
the $A_{V}$ for each YSO has been estimated using the NIR TCD as follows.
The interstellar extinction $A_{V}$ for a YSO lying in the `F' and `T' region of NIR TCD (cf. Fig. 7) has been obtained by
tracing them back to the intrinsic locus of TTSs. 
In the  case of other sources which do not lie in `F' and `T' regions, the $A_{V}$ of the YSO lying spatially near to the source is applied. 
Considering the uncertainties that might have gone into the estimation of $A_{V}$ value of each source, we used the estimated value $A_{V}$ of each individual source with a possible range of error as $A_{V}\pm$2 mag as input to the model. 
The distance range
is given as 3.0 to 3.6 kpc.
The error in NIR and MIR flux estimates due to possible uncertainties in the calibration, extinction, and intrinsic object variability was set as 10\%-15\%.

The physical parameters like accretion disk mass and accretion rate of probable YSOs constrained from
the SED fitting tool are given in Table 5. They have been obtained using the criterion $\chi^{2}$-$\chi^{2}_{min}$$\le$ 3N$_{data}$ as suggested by Robitaille et al. (2007), where $\chi^{2}_{min}$ is the goodness of fit parameter for the best fit model and N$_{data}$ is the number of input observational data points. The
SED fitting for two variables V10 an V11 is shown in Fig. 13 as examples.
 The tabulated parameters are obtained from the weighted mean and standard deviation of all the models that satisfy the above mentioned criterion, weighted by the inverse square of the $\chi^{2}$ of each model. The errors in accretion disk mass and accretion rate given in Table 5 are large because we are dealing with a large parameter space with a limited number of observational data points especially towards longer wavelengths.

\section{Characteristics of Variable stars}
\subsection {Non-periodic variables}
A few stars stars in our sample seem to show irregular brightness variations. 
Star V26 seems to have irregular photometric variations (Fig. 9 ). This star could be irregular variable and its period estimate
might be wrong.  
The light curve of V51 also shows irregular variations (See Fig. 5). On first intra-night observations, 
its brightness dropped up to $\sim$ 0.1 mag and within one and half hour its luminosity
appeared to reach its maximum brightness, while on second intra-night observations it showed no sign of brightness modulation and remained at its maximum brightness. 
If we consider it to be a short period variable, 
the first two intra-night observations yield a period of 
0.36 day, whereas the last two intra-night observations on 2008 October 29 and 2008 November 21 give a period of 0.61 day.
The entire data set for this star suggests a period of 
2.86 day. This indicates that star V51 is either pulsating with multiperiod or it might be
an eruptive variable.
As discussed in Section 4 this star could be a  B-type star.  
Since  B-type stars are found to have multiple pulsation periods, it  could be multiperiodic  B-type
star. Its absolute magnitude (0.47 mag) and intrinsic $B-V$ colour (-0.24 mag) put this star in the $H-R$ diagram in the 
location of $\gamma$ Cassiopeiae variables, which are irregular
variable having outflow of matter. The ($B-V)$ value of star has been taken from WEBDA\footnote[5]{http://www.univie.ac.at/webda/}.

\begin{figure*}
\includegraphics[width=14cm]{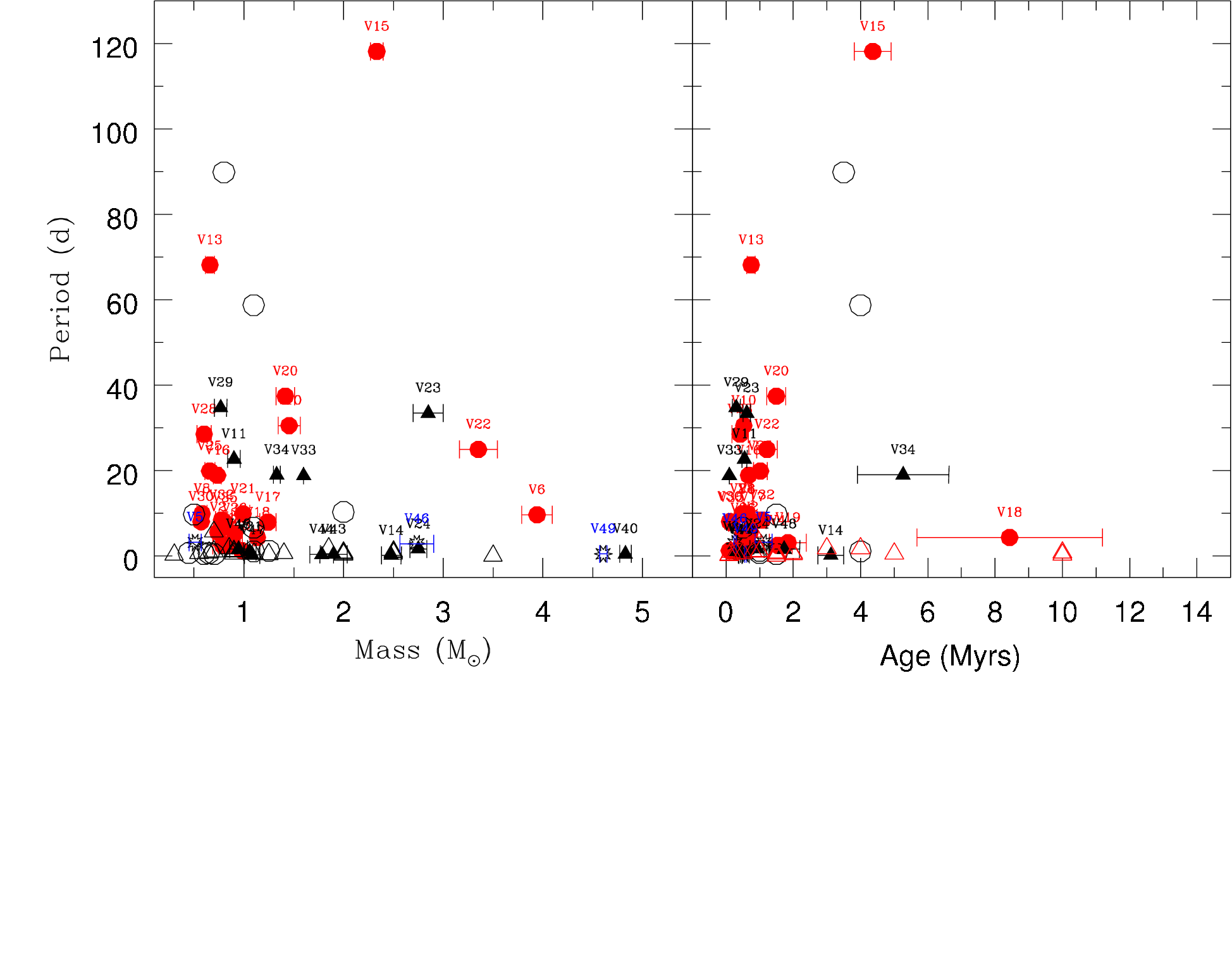}
\caption{Rotation period of TTSs as a function of mass and age. The symbols are same as in Fig. 8. Starred circles represent probable YSOs. One probable YSO V27 (P=220.93) is not plotted in the figure. Open circles and triangles represent data for Be 59 taken from Lata et. al. (2011).}
\end{figure*}
\begin{figure*}
\includegraphics[width=14cm]{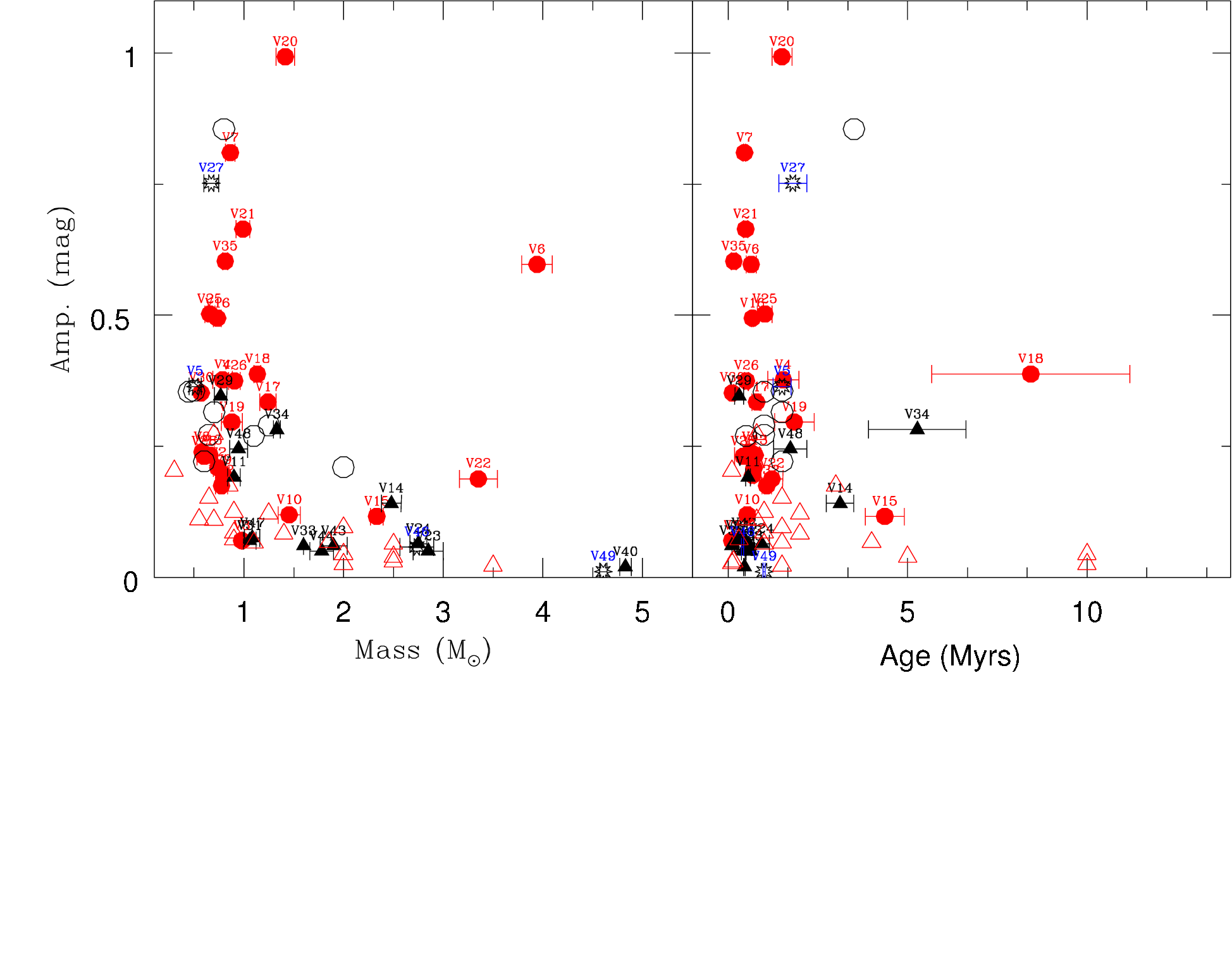}
\caption{Amplitude of TTSs as a function of mass and age. The symbols are same as in Fig. 8 and Fig. 14. Open circles and triangles represent data for Be 59 taken from Lata. et. al. (2011).} 
\end{figure*}

\subsection{Periodic variables}
The periodic variations in TTSs 
are believed to occur due to the axial rotation 
of a star with an inhomogeneous surface, having either hot or cool spots (Herbst et al. 1987, 1994). Herbst et al. (1994) identified 
three types of day to weeks timescale variability in TTSs. Type I variability, most often seen in WTT variables, is characterized 
by a smaller stellar flux variations (a few times 0.1 mag) and results from the rotation of a cool spotted photosphere. 
Type II variables have larger brightness variations (up to ∼2 mag), most often irregular but sometimes periodic and associated with 
short-lived accretion related hot spots at the stellar surface of CTT variables. The rare type III variations are characterized by 
luminosity dips lasting from a few days up to several months, which presumably result from circumstellar dust obscuration.

The majority of the PMS variables detected in the present study have masses $\lesssim$ 3 M$_{\odot}$ and these variables could be TTSs. The estimated periods of these probable TTSs are in the range of 0.15 to 118.09 day, with $\sim$ 75\% having periods $\le$ 20 day. 
The period estimates of CTTSs range from 0.36 day to 118.09 day, however majority ($\sim$ 75\%) have periods less than 20 day.  
Two CTTSs V13 and V15 are found to have longer periods (68.13 day and 118.09 day, respectively). The spectral energy distributions of V13 and V15 indicate 
IR excess, and H$\alpha$ emission in the case of V15, suggesting circumstellar disks. Edwards et al. (1993) found that 
the stars whose $H-K$ colours indicating the presence of circumstellar disks rotate slower than  whose $H-K$ 
colours indicate the absence of an accretion disk. The period estimates for WTTSs vary from 0.15 day to 34.73 day.

The amplitude in case of CTTSs has a range of 0.07 to 0.99 mag, while amplitude of WTTSs varies from 0.02 to 0.35 mag. 
This indicates that the brightness of CTTSs varies with larger amplitude in comparison to WTTSs. This result is 
 in agreement with that by Grankin et al. (2007, 2008) and by Lata et al. (2011). The larger amplitude in the case of CTTSs could be due to 
presence of hot spots on the stellar surface produced by accretion mechanism. Hot spots cover a small fraction of the stellar surface 
but with a high temperature causing larger amplitude of brightness variations (Carpenter et al. 2001). The smaller amplitude in WTTSs suggests 
dissipation of their circumstellar disks or these stars might have cool spots on their 
surface which are produced due to convection and differential rotation of star and magnetic field.

In the present study we have identified 19 short period variables. The variability characteristics for some of them are
discussed individually.
The location of V12 on the $U-B/B-V$ TCD and NIR TCD suggests 
an A-type MS star.
It shows persistent variability in its light curve. 
Its period (0.31 day) is 
similar to that of W UMa-type eclipsing binary. 
The shape of the light curve also resembles a W UMa-type variable star, hence
it could be A-type W UMa binary star. 

The star V14 (V=15.761 mag, mass=2.48 M${_\odot}$) is  
classified as a WTTS. It shows H$\alpha$ emission.
It is a periodic variable with amplitude of 0.16 mag with a period 
0.15 day. 
However its individual night observations 
on 2008 October 29 and 2010 November 27 give the period of 
0.15 and 0.17 day respectively. Its variability characteristics are rather similar to a
pulsating PMS star. Its absolute magnitude and intrinsic $(B-V)$ colour put it on the pulsation strip in the $H-R$ diagram where $\delta$ Scuti stars are expected to be located.
We feel that star V14 could be a PMS $\delta$ Scuti type star.
Several previous studies (e.g. Breger 1972; Kurtz \& Marang 1995; Marconi et al. 2000;
Donati et al. 1997; Zwintz et al. 2005, 2009; Zwintz \& Weiss 2006) have already reported PMS pulsators ($\gtrsim$ 2M$_{\odot}$)
in the instability region as they evolve to the MS.
A combination of mass, temperature and luminosity of these stars make them to pulsate (Zwintz et al. 2005, 2009).

The light curve of V36 indicates  
an eclipsing binary. It is a relatively faint star (V=17.958 mag) located at $\sim6.0 $ arcmin away from the cluster centre. 
Stars V37 and V38 are found to be located on the MS in the $V/V-I$ CMD.
Star V38 is a confirmed  B-type MS star with an emission-line spectrum (Zhang et al. 2008), whereas V37 could also be a
 B-type MS star according to Marco \& Negueruela (2002).  
Their variability characteristics like amplitude, period and shape of the light curve also suggest B-type
pulsating stars.

The locations of stars V39, V40, V42, V45, V49, V52 and V53 in the $V/V-I$ CMD (Fig. 8) indicate 
that these stars have masses $\gtrsim$3 M${_\odot}$. Marco \& Negueruela (2002)
have classified V40 as a Herbig Be star with a spectral type B0.5 IVe. The 
light curves of V39, V40 and V42 are rather similar, with periods 0.32 day, 0.48 day and 0.31 day and amplitudes 0.02, 0.20 and 0.02 mag, respectively.
The light curves of V49 shows a different behaviour from these three stars.
The location of V39 and V49 in the $J-H/H-K$ TCD lies in the region of
Herbig Ae/Be stars. The location of V39, V40 and V49 in $U-B/B-V$ TCD 
suggests that these star probably have excess in the $U$-band. We classify V39, V40, V42 and V49 as Herbig Be type stars.
The variability characteristics of V45, V52 and V53 show that these stars could be  B-type stars.

The period and shape of the light curve of star V41 (period=0.67 day and amp=0.04 mag) indicate that it could be a pulsating 
star similar to 
an RR Lyrae variable. 
RR Lyrae variables are similar to Cepheids, but less luminous and have periods of 0.5 to 1 day. 
 Its spectral type F0 III-IV (Marco \& Negueruela 2002) also suggests an RR Lyrae type variable.
The above discussion indicates that this star could not be young enough to be a part of the cluster NGC 1893. 
 
The location of V50 on the $J-H/H-K$, $U-B/B-V$ CC diagram and $V, V-I$ CMD
indicates that it could be a MS  B-type star.
The shape of the light curve and variability characteristics (period=0.58 day and amp=0.02 mag) 
put it in a category of  B-type pulsating stars. Its absolute magnitude
 and intrinsic $(B-V)$ colour 
lie in the region of $\beta$ Cep variable stars in the $H-R$ diagram. 
Its light curve is also similar to that of $\beta$ Cephei type
variable stars. The $\beta$ Cephei type
variable stars 
have periodical pulsations in the range of 0.1 to 0.6 day with an amplitude of 0.01 to 0.3 mag. 

\begin{figure*}
\includegraphics[width=14cm]{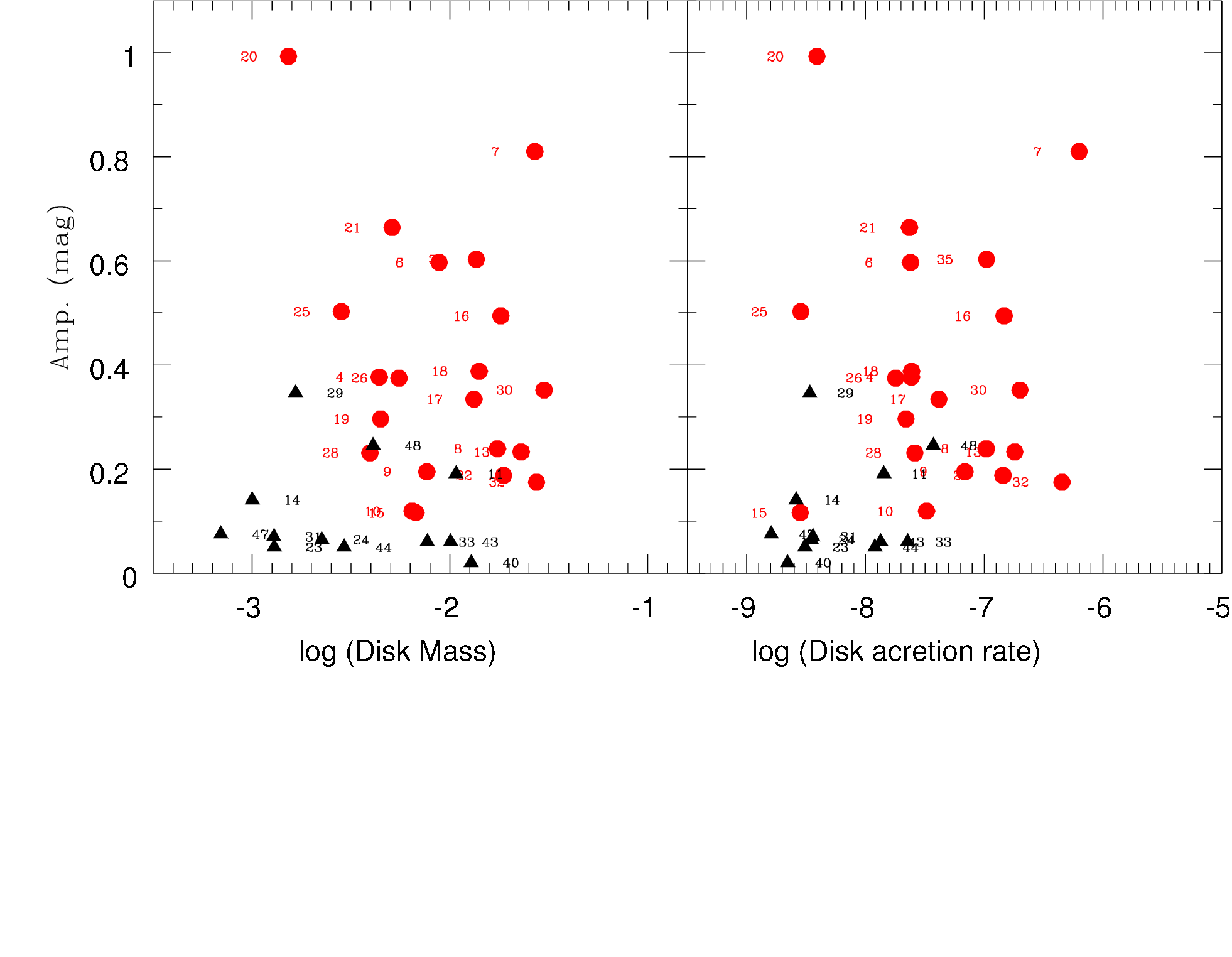}
\caption{Amplitude of TTSs as a function of disk mass and disk mass accretion rate.  The symbols are same as in Fig. 8.}
\end{figure*}

To understand the correlation of mass and age of the TTSs on the rotation period we plot
the rotation period as a function of mass and age in Fig. 14. 
To 
increase the sample we have also included data for Be 59 from Lata et al. 
(2011) because this cluster has similar environment to the NGC 1893, and moreover the same technique has been used to identify variable stars and determine their physical properties. Although there is a large scatter, the stars having masses $\gtrsim$2$M_{\odot}$ are found to be fast rotators.  Similarly, the stars having ages $\gtrsim$3 Myr seem to be fast rotators. This result is  
compatible with the disc locking model in which it is expected that the population of stars with disc rotate slowly than those without disc (Edwards et al. 1993; Herbst et al. 2000; Littlefair et al. 2005). 
In a recent study Littlefair et al. (2011) pointed out ``if the location of a star in the CMD  is  interpreted as being 
due to genuine age spreads within a cluster, then the implication is that the youngest stars in the cluster 
(those with the largest moments of inertia and highest likelihood of ongoing accretion) are the most rapidly rotating. 
Such a result is in conflict with the existing picture of angular momentum evolution in young stars, where 
the stars are braked effectively by their accretion discs until the disc disperses''.  Alternatively, they argued that 
the location of the stars in the CMD is not primarily a function of age, but of accretion history. They discussed that 
this hypothesis can, in principle, explain the observed correlation  between rotation rate and location of 
the star in the CMD. It is worthwhile to point out that the study by Littlefair et al. (2011) assumes that the 
stars in the  mass range 0.4$\leq M/M_{\odot} <$1  have little dependence between rotation rate and stellar mas. However, in the present study we find that relatively massive stars in the mass range 0.5$\leq M/M_{\odot} <$1.3 have faster rotation.

Fig. 15 reveals that amplitude of TTSs variability is correlated with the mass (left panel)
and age (right panel) in the sense that amplitude decreases with increase in mass as well as age of variable star. 
The star V6 does not seem to follow the mass-amplitude trend. This star
 has a large IR excess and shows a large variation in the amplitude (0.10 and 0.29 mag: Zhang et al. 2008.; 0.60 mag: present work).
 It could be either a star whose brightness variation is
supposed to be produced by obscuration due to dust clumps or clouds as in the case of Type III
variability characteristics suggested by Herbst et al. (1994) or it might be an
eclipsing
binary.
The stars V18 (Class II source) and V34 (Class III source) have ages $>$5 Myr and do not follow the general decreasing trend in amplitude-age distribution (Fig. 15, right
panel).
The star V18 is undoubtedly a Class II source as it shows NIR excess.
Its amplitude is $\sim$ 0.40 mag. If its $V$ magnitude given in Table 2 is at minimum,
its age will be overestimated on the basis of $V/V-I$ CMD. Similarly the
star V34 has amplitude $\sim$ 0.3 mag. Its
 mean $V$ magnitude could yield an age of $\sim$3 Myr. Baraffe et al. (2009) have shown that the higher age of young low mass stars could be 
results 
of episodic accretion which produces a luminosity spread in the HR diagram at ages of a few Myr. 

The decrease in amplitude could be due to the dispersal of the disk. This
result further supports the notion, as obtained in our previous study of Be 59 (Lata
et al. 2011) that the disk dispersal mechanism is less efficient for relatively
low mass stars. Fig. 15 (right panel) manifests that significant amount of
the disks is dispersed by $\lesssim$ 5 Myr. This result is in accordance 
with the result obtained by Haisch et al. (2001).

Fig. 16 plots amplitude as a function of disk mass and disk accretion rate. We did not find any correlation between amplitude and disk mass as well as disk accretion rate.
\section{Summary}
The paper presents time series photometry of 53 variable stars identified in the cluster NGC 1893 region.
The probable members associated with the cluster are identified on the basis of spectral energy distribution,
location in the $V/V-I$ CMD and $J-H/H-K$ TCD.
Forty three variables are found to be probable PMS stars. The majority of these sources 
have ages $\lesssim$5 Myr and masses in the
range of 
0.5 $\lesssim$ $M/M_{\odot}$$\lesssim 4$ and hence should be TTSs.
The rotation periods of majority of TTSs ranges from 0.1 to 20 day.
The brightness of CTTSs varies with larger amplitude in comparison to that of WTTSs.
Both the period  and amplitude of variability of TTSs decrease with increasing mass.
This confirms
our earlier finding that decrease in amplitude of variability in relatively massive stars could be due to the dispersal of
circumstellar disk. This result supports our earlier result (Lata et al. 2011) that mechanism
of disk dispersal operates less efficiently for relatively low mass stars.

\section{Acknowledgments}
Authors are very thankful to the anonymous referee for a critical reading of the paper and useful scientific suggestions that improved scientific content of the paper.
Part of this work was carried out by AKP during his visit to National Central
University (Taiwan) under India-Taiwan collaborative program. AKP is thankful
to the GITA, DST (India) and NSC (Taiwan) for the financial help.

\bibliographystyle{mn2e}

\end{document}